\documentclass[%
 reprint,
 amsmath,amssymb,
 aps,
]{revtex4-2}

\usepackage{graphicx}% Include figure files
\usepackage{dcolumn}% Align table columns on decimal point
\usepackage{bm}% bold math
\usepackage{hyperref}% add hypertext capabilities
\usepackage{array}
\usepackage{xcolor}
\usepackage[normalem]{ulem}
\usepackage{orcidlink}

\begin{document}

\preprint{APS/123-QED}

\title{Half-wave-plate non idealities propagated to component separated CMB $B$-modes }% Force line breaks with \\
%\thanks{A footnote to the article title} 

\author{Ema Tsang King Sang$^1$ 
\orcidlink{0009-0001-6108-9518}}
\email{tsang@apc.in2p3.fr}
\author{Josquin Errard$^1$
\orcidlink{0000-0002-1419-0031}}
\email{josquin@apc.in2p3.fr}
\author{Simon Biquard$^{2,1}$\,\orcidlink{0000-0002-1493-2963}}
\author{Pierre Chanial$^1$
\orcidlink{0000-0003-1753-524X}}
\author{Wassim Kabalan$^1$
\orcidlink{0000-0003-2651-0314}}
\author{Wuhyun Sohn$^1$
\orcidlink{0000-0002-6039-8247}}

\author{Radek Stompor$^1$
\orcidlink{0000-0002-9777-3813}}

\affiliation{$^1$Universit\'e Paris Cit\'e, CNRS, Astroparticule et Cosmologie, F-75013 Paris, France}
\affiliation{$^2$Jodrell Bank Centre for Astrophysics, University of Manchester, Oxford Road, Manchester, England M13 9PL, United Kingdom}

\date{\today}% It is always \today, today,

\begin{abstract}
We assess the impact of non-ideal, continuously rotating half-wave plates (HWPs) on cosmic microwave background (CMB) polarization measurements targeting large angular scale signal. Such hardware solutions are used in or planned for multiple modern CMB efforts, both ground-based, for instance, small aperture  telescopes of \textsc{Simons Observatory} or satellite borne, such as \textsc{LiteBIRD}. Using a frequency-dependent parametric model based on the Mueller matrix formalism, we characterize the induced mixing of Stokes parameters. Through end-to-end simulations, we propagate these effects from time-ordered data to cosmology via map-making and component-separation stages, quantifying their impact on the $B$-modes power spectrum and the tensor-to-scalar ratio, $r$. Our analysis shows that neglecting the frequency dependence of a three-layer HWP gives rise to significant polarization leakage, biases foreground spectral parameters, and thus leads to residual contamination in the recovered CMB maps. To mitigate these effects, we investigate multiple analysis strategies progressively incorporating a more complete description of the instrumental response. At the map-making level, this requires generalizing the standard pointing matrix, to account not only for the scanning strategy but also for the full time- and frequency-dependent instrumental response. We find that two standard HWP models, referred to as effective and stack HWP models, reduce the biases only down to $r \sim 10^{-2}$, however a more advanced approach based on a generalization of both map-making and component separation procedures, implemented using \texttt{JAX}, can suppress it down to $r \sim 7 \times 10^{-4}$. Finally, we extend this approach to a time-domain component-separation framework, enabling a statistically consistent treatment of instrumental response in the presence of time-domain features such correlated noise. 
We demonstrate its feasibility and validate it by performing a full end-to-end analysis, recovering results in good agreement with the map-based ones. This sets the stage for the full exploitation of this approach’s capability in the future.
\end{abstract}
            
\maketitle

%\tableofcontents

\section{Introduction} \label{sec:intro}

Over the past decade, a significant focus of Cosmic Microwave Background (CMB) experiments has been on the detection of primordial $B$-modes polarization. These faint polarization patterns in the CMB anisotropies are a key signature of primordial gravitational waves, which are predicted by theories of cosmic inflation~\cite{Kamionkowski_2016}.

Achieving this requires overcoming numerous challenges due to the expected small amplitude of primordial $B$-modes signal compared to observational noise, astrophysical and instrumental systematics, and environmental contamination, e.g.~\cite{Errard_2016}. As the targeted signal is predominantly on the large angular scales, the effects present on large angular or long temporal scales are of particular relevance. 
Among the former ones, the presence of emissions from other astrophysical sources, collectively known as foregrounds \cite{Krachmalnicoff_2018} is particularly challenging. These include radiation from our own galaxy —such as synchrotron and thermal dust emissions— as well as signals from extragalactic sources. Foregrounds exhibit spatial and frequency-dependent variations across the sky, and their accurate identification and removal are therefore essential. To tackle this issue, CMB experiments rely on multi-frequency observations and component separation techniques, which leverage the distinct frequency and spatial signatures of the CMB and foreground emissions to disentangle and isolate the cosmological signal~\cite{Delabrouille_2008}. 

The long-timescale effects can be mitigated via efficient modulation of the incoming signals, be that through an appropriately adapted scanning strategy or suitable hardware solutions, incorporating polarization modulators.
The modulators shift the sky signal to higher frequencies in the detector timestream, away from the low temporal effects. They are particularly, relevant for ground-based experiments as they allow for suppressing atmospheric fluctuations. They can also mitigate the detector $1/f$ noise, and differential systematic effects, and therefore can be of interest for future satellite missions, as it is indeed the case of \textsc{LiteBIRD}.
Continuously rotating Half-Wave Plates (HWPs) have become particularly popular and have been widely used in experiments such as 
\textsc{Polarbear}~\cite{hill2016design}, 
\textsc{SPIDER}~\cite{bryan2014half}, 
\textsc{ABS}~\cite{Kusaka_2014}, 
\textsc{MAXIPOL}~\cite{Johnson_2007}, 
and \textsc{EBEX}~\cite{klein2011cryogenic}, as well as in the ongoing 
\textsc{Simons Observatory} Small Aperture Telescopes~\cite{yamada2024simons,sugiyama2024simons}, becoming a critical component to enhance sensitivity and control diffrential systematic effects ~\cite{Takakura_2017}.

A rotating HWP modulates the incoming linear polarization at four times the mechanical rotation frequency of the plate. 
For a typical rotation frequency $f_{\rm HWP} = 2$~Hz, the polarized signal is modulated at $4 f_{\rm HWP} = 8$~Hz, which lies well above the knee frequency of the detector $1/f$ noise, typically $f_{\rm knee} \approx 1$--$2$~Hz for ground-based observatories. 
Given the particular time-modulation of the polarized signal, the HWP also mitigates systematic errors that would otherwise result
from the differential response of detectors sensitive to orthogonal polarizations (source of intensity to polarization leakage).
However, their practical implementation is not without limitations. In particular, real HWPs 
depart from the idealized case: the imposed phase delay corresponds to a half-wave only at a single 
frequency, while instruments operate over finite bandwidths. Multi-layer HWP stacks (sapphire or meta-material) are thus employed to broaden the effective bandwidth~\cite{pancharatnam1955}, but at the cost of introducing additional 
systematic effects that must be carefully characterized and mitigated \cite{matsumura2014,Giardiello_2023,Monelli_2023,patanchon2023,raffuzzi2026, duivenvoorden2021probing}.

We study the impact of the frequency dependence of sapphire HWP on the analysis of data from modern, multi-frequency CMB experiments, which need to separate Galactic contributions in order to isolate the signal of cosmological interest. We first present the instrumental framework used to model and simulate the HWP (Sec.~\ref{sec:instmod}) 
and then introduce the end-to-end pipeline developed to quantify the impact of HWP 
non-idealities (Sec.~\ref{sec:pipeline}), focusing on frequency-dependent effects. We apply 
this methodology to a Stage~3 CMB experiment, exemplified by a \textsc{Simons Observatory}-like
configuration, to a CMB-only sky in Sec.~\ref{sec:appcmb} before adding foreground contribution in Sec.~\ref{sec:appdust}. Although we focus on the specific challenges relevant to ground-based observations, the formalism is general and can be applied to other HWP implementations. In particular, it applies to any HWP whose Mueller matrix is fixed in the instrument frame and whose modulation arises solely from its rotation. Alternative designs, such as metamaterial HWPs, may modify the quantitative frequency-dependent response of the system, but do not change the structure of the modeling framework under this assumption.\\

This work describes and capitalizes on the tools and techniques developed in the context of the \textsc{SciPol} project \footnote{\href{https://scipol.in2p3.fr/}{\texttt{https://scipol.in2p3.fr/}}} 
(Science from the large scale cosmic microwave background polarization structure). It proposes a methodology based on an innovative joint analysis of instrumental and foreground effects. We have been developing \texttt{FURAX}~\footnote{\href{https://github.com/CMBSciPol/furax}~\cite{Chanial2026}{\texttt{https://github.com/CMBSciPol/furax}}}(Framework for Unified and Robust data Analysis with \texttt{JAX}), an open-source \texttt{Python} library that provides building blocks to construct instrument and noise models in a modular way. The idea is to create a framework to mathematically represent the analysis pipeline of a CMB experiment and carry out the heavy computations it involves in an efficient way by leveraging Just-In-Time compilation and GPU capabilities. It is a flexible framework, able to use generalized pointing matrix that could include the scaling of sky components (including atmosphere) as well as the instrumental response, that includes maximum-likelihood map-making and parametrized component separation (\cite{Chanial2026, Kabalan2025, Beringue2025, Sohn2026}). 

\section{Optical chain model} \label{sec:instmod}
In this section, we present the mathematical formalism required to incorporate a multi-layer HWP into the data model of a typical CMB polarization instrument. The HWP is assumed to be the first optical element 
in the chain. We adopt the Mueller matrix formalism, commonly used to describe polarized components. The Mueller matrix formalism is directly 
related to the Jones calculus, from which it can be derived for non-depolarizing optical elements. For incoming radiation characterized by the Stokes parameters $\mathbf{s} = (\mathrm{I}, \mathrm{Q}, \mathrm{U}, \mathrm{V})$, the Mueller matrix $\mathbf{M}$ of a polarizing element specifies how the state of $\mathbf{s}$ is transformed as the signal passes through the element with the linear relation:

\begin{equation}
    \mathbf{s}_{\rm{out}} = \mathbf{M} \, \mathbf{s}_{\rm{in}}
\end{equation}
where $\mathbf{M}$ is a (4x4) operator. We will assume in the following that the incoming $\rm{V}$-polarization is zero. 
This is well motivated for CMB observations, since the CMB is not expected 
to produce circular polarization at any detectable level, and Galactic 
foregrounds such as thermal dust and synchrotron emission are likewise 
consistent with vanishing $\mathrm{V}$ to the precision relevant here~\cite{Kamionkowski_2016}. We nonetheless retain the full $4\times4$ dimensions of the Mueller matrix 
operators throughout, as the non-ideal HWP can convert circular to linear 
polarization ($\rm{V} \rightarrow \rm{QU}$).

In practice, many experiments employ multi-layer stacks in order to increase the bandwidth of the HWP ~\cite{pancharatnam1955}. For instance, 
the \textsc{Simons Observatory} (SO) has deployed a three-layer sapphire stack on its medium frequency  (MF: $90/150$GHz) and ultra high frequency (UHF: $220/280$GHz) telescopes. In such a case, the Mueller matrix of the HWP depends on frequency $\nu$, incidence angle $i$, and a set of instrumental parameters 
$\gamma = (\theta, \xi)$ describing respectively the layer thickness and the relative orientation $\xi$ of the middle plate. 
Additional parameters can be included in $\gamma$ to describe more complex configurations without modifying the formalism.
\noindent For a detector of polarization angle $\alpha_t$ observing through a rotating HWP at angle $\varphi_t$, the optical chain is described as 

\begin{equation}
    \mathbf{M}(t) = \mathbf{M}_{\rm det}\,
    \mathbf{R}(-2\varphi_t)\,
    \mathbf{H}(\gamma)\,
    \mathbf{R}(2\varphi_t)\,
    \mathbf{R}(2\alpha_t).
    \label{eq:Mopticchain}
\end{equation}

\noindent \noindent where $\varphi_t$ is the rotation angle of the HWP, 
$\alpha_t$ the detector angle, with respect to the sky rest frame, $\mathbf{H}$ the HWP Mueller matrix, $\mathbf{R}$ is a rotation matrix around the optical axis, 
\begin{equation}
    \mathbf{R}(2\theta) =
    \begin{bmatrix}
        1 & 0 & 0 & 0 \\
        0 & \cos 2\theta & -\sin 2\theta & 0 \\
        0 & \sin 2\theta & \cos 2\theta & 0 \\
        0 & 0 & 0 & 1
    \end{bmatrix}.
\end{equation}

\noindent and  $\mathbf{M}_{\rm det}$ the detector modeled as a linear polarizer,
\begin{equation}
    \mathbf{M}_{\rm det} =
    \frac{1}{2}
    \begin{bmatrix}
        1 & 1 & 0 & 0 \\
        1 & 1 & 0 & 0 \\
        0 & 0 & 0 & 0 \\
        0 & 0 & 0 & 0
    \end{bmatrix}.
    \label{eq:Mdet}
\end{equation}

In CMB experiments, detectors measure the incoming polarized signal by projecting the full Stokes vector onto the detector’s sensitivity axis. This measurement is captured by the first row of the optical chain’s Mueller matrix product, yielding the observed time-ordered data (TOD) as a linear combination of the Stokes parameters:

\begin{align}
    \mathbf{d}_t &= \mathbf{M}_{00}(t) \, \mathrm{I}_t + \mathbf{M}_{01}(t) \, \mathrm{Q}_t + \mathbf{M}_{02}(t) \, \mathrm{U}_t, \label{eq:tod} \\
    \mathbf{M}(t) &= \prod_{i \in \text{elmt}} \mathbf{M}_i\,(t), \nonumber
\end{align}
where each $\mathbf{M}_i$ is the Mueller matrix of the i-th element of the optical chain, as defined in Eq.~\ref{eq:Mopticchain}

\subsection{Monochromatic HWP} \label{subsec:monohwp}

When the optical chain includes a rotating idealmonochromatic HWP,
\begin{equation}
    \mathbf{M}^{\mathrm{mono}}(t) = \mathbf{M}_{\rm det}\,
    \mathbf{R}(-2\varphi_t)\,
    \mathbf{H}_{\mathrm{mono}}\,
    \mathbf{R}(2\varphi_t)\,
    \mathbf{R}(2\alpha_t).
    \label{eq:Mopticchainid}
\end{equation}

\noindent the Mueller matrix product simplifies and the resulting time-ordered data (TOD) takes the characteristic form \cite{Johnson_2007}:
\begin{equation}
\mathbf{d}_t = \mathrm{I}_t + \cos(4\varphi_t+2\alpha_t)\,\mathrm{Q}_t + \sin(4\varphi_t+2\alpha_t) \,\mathrm{U}_t,
\label{eq:idtod}
\end{equation}
\noindent where $\mathrm{I}_t$, $\mathrm{Q}_t$, $\mathrm{U}_t$ the incoming sky Stokes parameters. We adopt the \texttt{IAU}~\footnote{\url{https://lambda.gsfc.nasa.gov/product/about/pol_convention.html}} convention throughout this work. 
%\simon{I think this is the correct form regardless of the convention (as long as the angles are consistently measured from the same reference, which is different for \texttt{COSMO} and \texttt{IAU}.} 
The Mueller matrix of an ideal monochromatic HWP (phase delay $\pi$) is written as 
\begin{equation}
    \mathbf{H}_{\rm mono} =
    \begin{bmatrix}
        1 & 0 & 0 & 0 \\
        0 & 1 & 0 & 0 \\
        0 & 0 & -1 & 0 \\
        0 & 0 & 0 & -1
    \end{bmatrix},
    \label{eq:idhwp}
\end{equation}

\noindent In each measurement, the three Stokes parameters are combined. Under the monochromatic assumption, these parameters can be disentangled due to the distinct time-dependent modulation imposed on each one, see Eq.~\ref{eq:idtod}. This is what the so-called ``demodulation'' capitalizes within usual CMB analysis pipelines.

\subsection{Stacked HWP} \label{subsec:multihwp}

CMB experiments measure the incoming signal integrated across finite frequency 
bands, so the monochromatic approximation of Eqs.~\ref{eq:idtod} - \ref{eq:idhwp} is incorrect. %\simon{incorrect? it may be sufficient depending on the science case...}. 
Indeed, for polychromatic light, the Mueller matrix of a birefringent plate can be expressed as
\begin{equation}
    \mathbf{H}_{\rm layer}(\delta) =
    \begin{bmatrix}
        1 & 0 & 0 & 0 \\
        0 & 1 & 0 & 0 \\
        0 & 0 & \cos\delta & -\sin\delta \\
        0 & 0 & \sin\delta & \cos\delta
    \end{bmatrix},
    \label{eq:layer}
\end{equation}
where the induced phase delay $\delta$ depends on the frequency $\nu$, the 
the thickness $\theta_{\rm th}$, the ordinary and extraordinary refractive 
indices of the material, the angle of incidence $i$, and the angle $\chi$ 
between the plate's optical axis and the optical axis of the instrument.

\begin{align}
    \delta &= \frac{2\pi \, \theta_{\rm th} \, \nu}{c}\, \lvert n(i) - n_o \rvert, 
    \label{eq:delta} \\
    n(i) & \equiv n_e \, \sqrt{1 + \big(n_e^{-2} - n_o^{-2}\big) 
    \sin^{2} i \cos^{2} \chi }.
    \label{eq:neff}
\end{align}
Here $n_e$ and $n_o$ denote respectively the extraordinary and ordinary refractive 
indices of the plate. Following the SO-like 3-layer configuration, the stacked HWP Mueller matrix can be written as
\begin{equation}
    \mathbf{H}_{\rm stack}(\nu, i, \gamma) =
    \mathbf{H}_{\rm layer}\, \mathbf{R}(-2\xi)\,
    \mathbf{H}_{\rm layer}\, \mathbf{R}(2\xi)\,
    \mathbf{H}_{\rm layer}.
    \label{eq:Mstack}
\end{equation} 

\noindent The complete Mueller matrix accounting on the HWP rotation is then given by Eq.~\ref{eq:Mopticchain} and is denoted hereafter as $\mathbf{M}^{\textrm{stack}}$, 

\begin{equation}
    \mathbf{M}^{\mathrm{stack}}(t) = \mathbf{M}_{\rm det}\,
    \mathbf{R}(-2\varphi_t)\,
    \mathbf{H}_{\mathrm{stack}}(\nu, i, \gamma)\,
    \mathbf{R}(2\varphi_t)\,
    \mathbf{R}(2\alpha_t).
    \label{eq:Mopticchainstack}
\end{equation}

\subsection{Non-ideal HWP} \label{subsec:nonidealhwp}

\begin{figure}[ht!]
    \centering
    \includegraphics[width=\columnwidth]{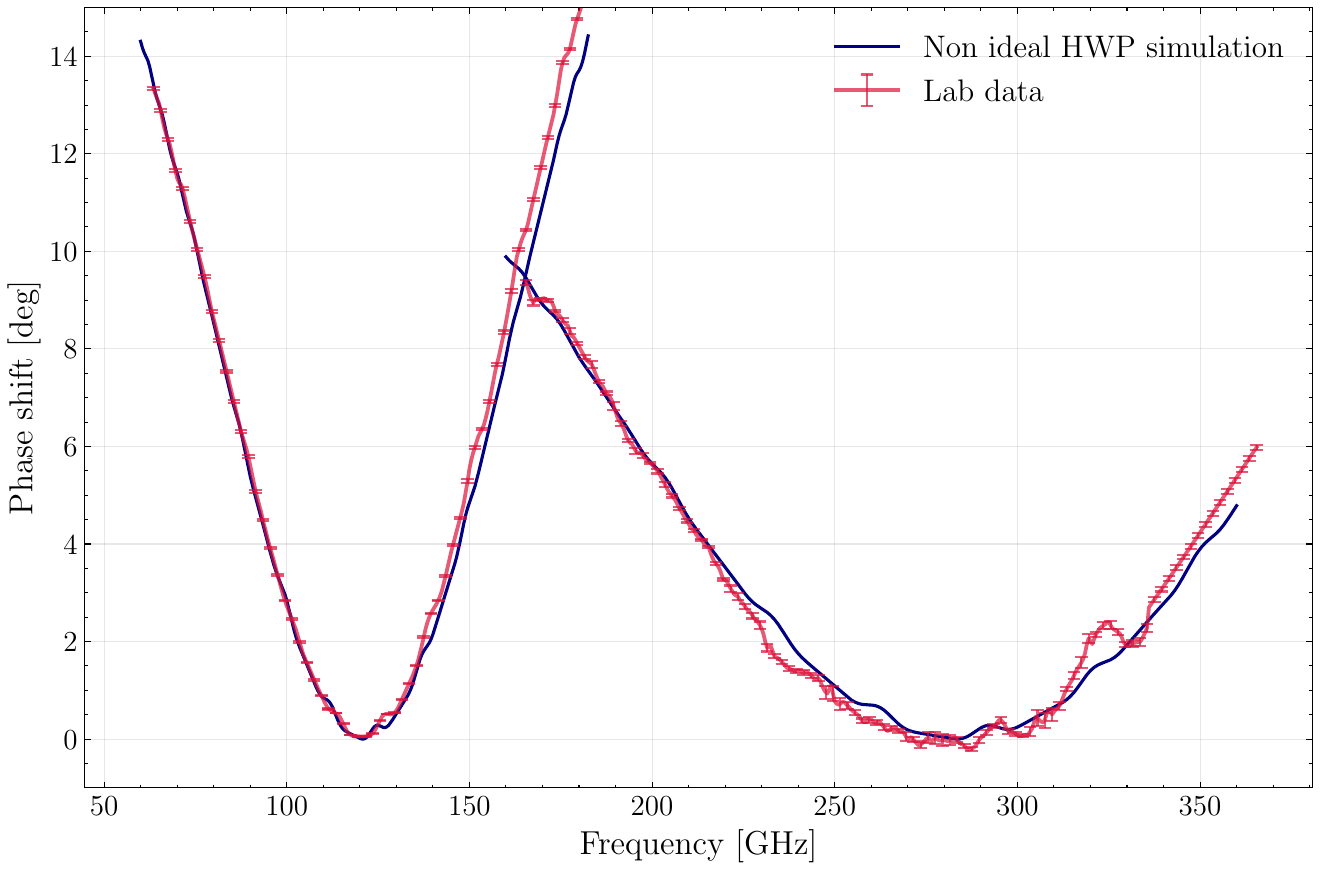}
    \caption{Phase shift of the HWP as a function of frequency. The blue line shows the prediction of the non-ideal HWP optical model, while the red curve shows laboratory measurements with $1\sigma$ uncertainties from \cite{sugiyama2024simons}. The measurements span the MF and UHF frequency ranges and show good agreement with the model across most of the band. A constant offset between the curves is applied since the phase shift is defined only up to an arbitrary additive constant.} 
    \label{fig:labdata}
\end{figure}

Although ideal and stacked HWP models provide valuable insights into the modulation of polarized light, real applications require accounting for additional physical effects. 
To better capture the reaslism of the HWP used in the telescope, we introduce a model, $\mathbf{H}_{\rm nonid}$, which explicitly accounts for complex optical effects such as anti-reflection coatings~\cite{sakaguri2024antireflection} that are not included in the Jones-based description. This model is derived using the transfer matrix method~\cite{essinger2013transfer}, which computes the transmission and reflection properties of light by enforcing the continuity of electromagnetic field across material boundaries, as prescribed by Maxwell's equations. This approach is able to provide an accurate description of polarized light propagation through the HWP, allowing more precise simulations of the observed signal. It has indeed been shown to provide a good representation of laboratory measurements, ensuring its reliability for simulating realistic observational conditions ~\cite{sakaguri2024antireflection, sugiyama2024simons}, Fig.~\ref{fig:labdata}. We stress that all of these effects are direct consequences of classical electromagnetic theory and are to be expected in all realistic scenarios~\cite{Moncelsi_2013, Monelli_2023, Essinger_Hileman_2016}.

\begin{figure}[ht!]
    \centering
    \includegraphics[width=\columnwidth]{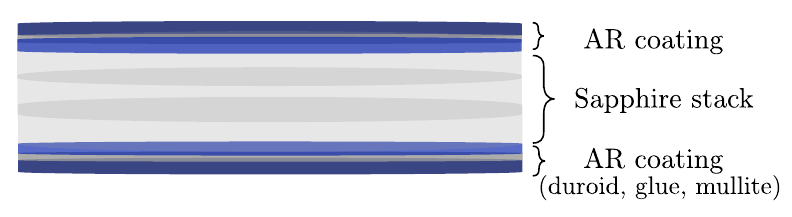}
    \caption{HWP configuration consisting of a three-layer sapphire stack with 
    Mullite and Duroid anti-reflection coating, following \cite{sakaguri2024antireflection} 
    but omitting the alumina coating.}
    \label{fig:hwpconfig}
\end{figure}

In this study, we adopt a configuration consisting of a three-layer sapphire HWP with a Duroid--Mullite anti-reflection coating (Fig.~\ref{fig:hwpconfig}), a summary of the instrumental parameters used in this work is given in Tab.~\ref{tab:instparams}.

\begin{figure*}[ht!]
    \centering
    \includegraphics[width=\textwidth]{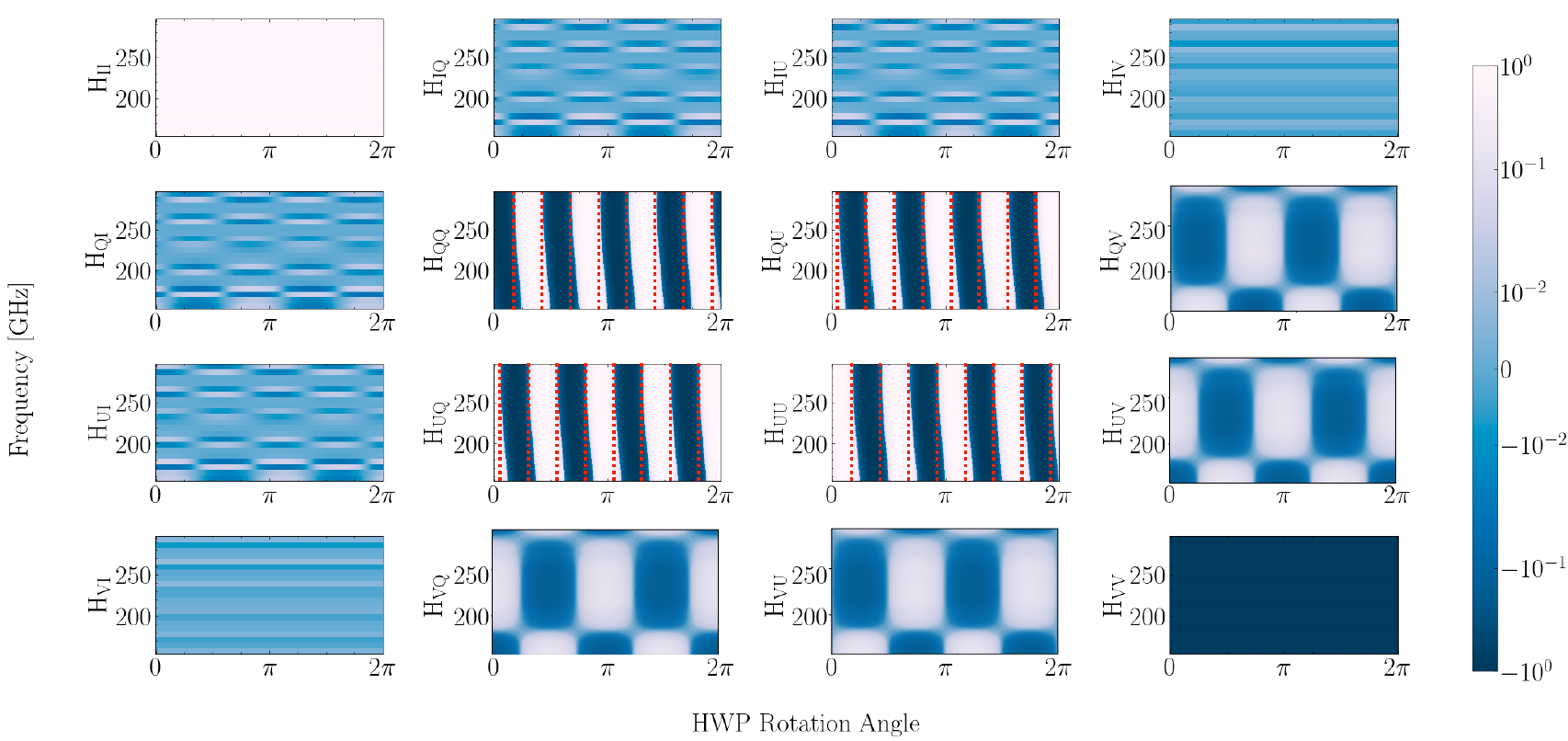}
    \caption{$4\times4$ rotating Mueller matrix $\mathbf{H}(\varphi_t)$ of a 3-layer sapphire HWP  computed with the transfer matrix method~\cite{essinger2013transfer}. 
The matrix elements are shown as a function of HWP rotation angle 
$\varphi_t$ and frequency, with the ideal HWP values overlaid as red dashed 
lines (frequency-independent). We observe a frequency-dependent phase 
shift in the central $2\times2$ $(\rm{Q,U})$ block, along with reduced 
modulation efficiency compared to the ideal case. The off-diagonal 
elements reveal an $\mathrm{I} \to \mathrm{QU}$ leakage of $0.05\%$ 
and a $\mathrm{V} \to \mathrm{QU}$ leakage, both predominantly modulated 
at $2f$.}
    \label{fig:2dhwp}
\end{figure*}

Some elements of the \(4\times 4\) Mueller matrix of the non-ideal HWP are shown in Fig.~\ref{fig:2dhwp} for $220-280$ GHz and normal incidence, $i=0\,\deg$. They exhibit a frequency-dependent phase shift of the polarization vector modulation in the central \(2\times 2\) ($\mathrm{Q}$,$\mathrm{U}$) block, reduced modulation efficiency compared to the ideal case, an intensity-to-polarization ($\mathrm{I}$ $\to$ $\mathrm{QU}$) leakage at the level of \(0.05\%\), and a circular-to-linear polarization ($\mathrm{V}$ $\to$ $\mathrm{QU}$) leakage modulated at \(2f\), in agreement with observations by~\cite{patanchon2023}.

\begin{figure*}[ht!]
    \centering
    \includegraphics[width=\textwidth]{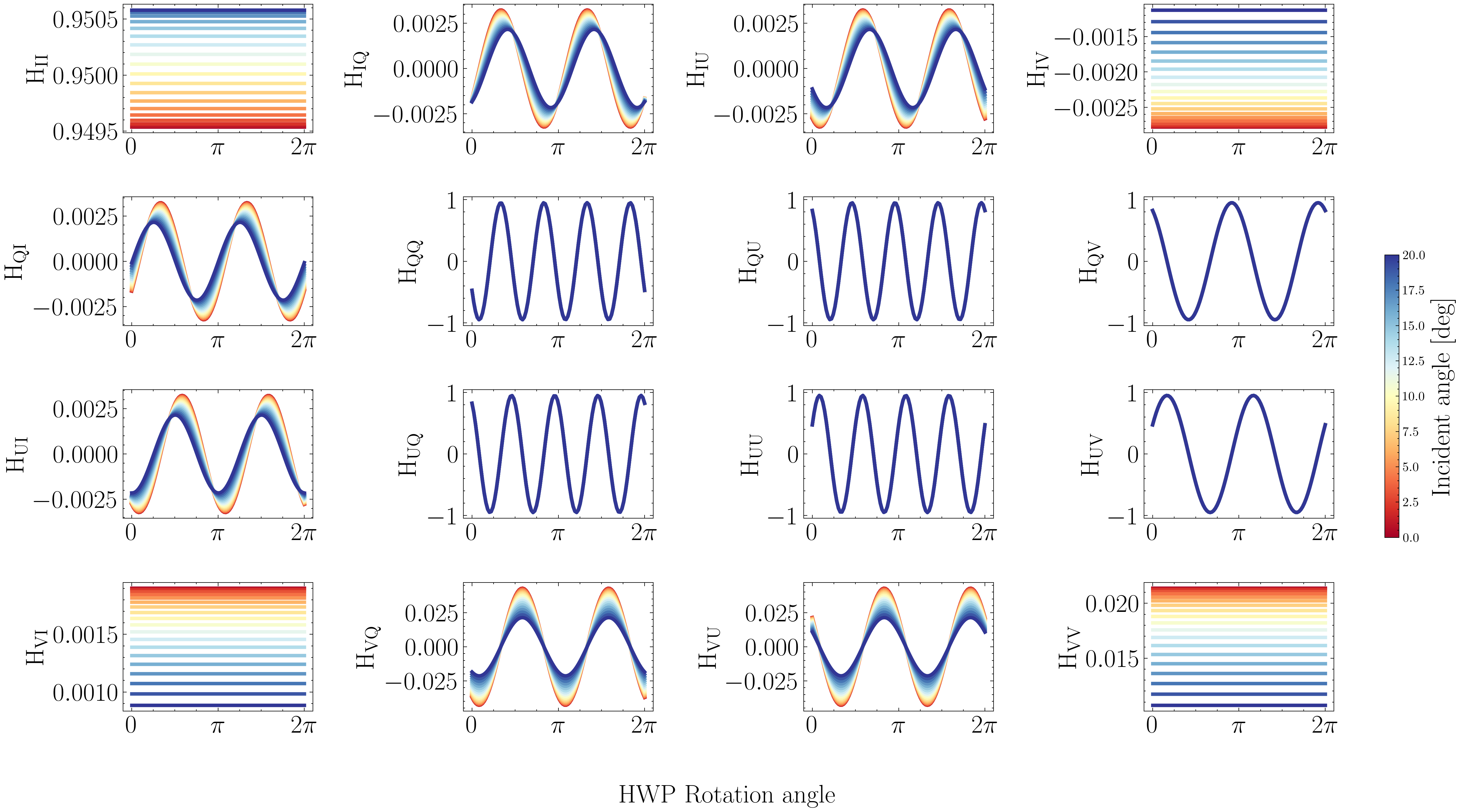}
    \caption{$4\times4$ rotating Mueller matrix $\mathbf{H}(\varphi_t)$ of a 3-layer sapphire HWP  computed with the transfer matrix method~\cite{essinger2013transfer} for different incident angles corresponding to a field-of-view of 20deg}
    \label{fig:incidenceangle}
\end{figure*}

Furthermore, this approach allows us to explicitly include key physical parameters, such as the thickness of each layer and the angle of incidence. As illustrated in Fig.~\ref{fig:incidenceangle}, the modulation patterns of the Mueller matrix elements are shown as a function of the HWP rotation angle for various incident angles up to half of a 35-deg field-of-view, represented by the color gradient. The characteristic 2$f$ and 4$f$ peaks are clearly visible in the plots, with the 2$f$ peaks exhibiting noticeable variations as a function of the incident angle. Additionally, the 4$f$ peak, while less sensitive to the incident angle at single frequency but shows significant variation with frequency. 
These non idealities result into HWP-synchronous-signal (HWPSS) that is not to be neglected in the case of non-ideal polarizer as discussed in~\cite{salatino2018}.

The frequency-dependent HWP effects are incorporated into the simulated time-ordered data through the following data model \cite{El_Bouhargani_2022} at the given frequency $\nu$:
\begin{equation}
    \mathbf{d}^{\rm non\, \rm id}_{t}(\nu_c) = \int \mathrm{d}\nu \, \boldsymbol{\mathcal{B}}(\nu_c, \nu)\,
    \mathbf{P}^{\rm input}_{{\rm non\, \rm id},(t,p)} \, \mathbf{s}_p(\nu),
    \label{eq:datamodel1}
\end{equation}
where $\boldsymbol{\mathcal{B}}(\nu_c,\nu)$ denotes the instrumental bandpass centered at $\nu_c$, and $\mathbf{s}_p(\nu)$ is the sky signal evaluated at pixel $p$ and frequency $\nu$. The pointing matrix $\mathbf{P}^{\rm input}_{\rm non id}$ encodes the scanning strategy, detector orientation, and instrumental response through the Mueller matrix of the optical chain defined in Eq.~\ref{eq:Mopticchain}. The index ${\rm non\, \rm id}$ labels the HWP model adopted in the simulations. Unless stated otherwise, the input TODs (left-hand side in Eq.~\ref{eq:datamodel1}) are generated using the non-ideal HWP model introduced above. Eq.~\ref{eq:datamodel1} does not impose a specific normalization of the band-integrated signal. In this work, we assume that the bandpassess as included on the simulation and analysis stages, are perfectly known, hence the normalization is irrelevant and we apply no correction in the simulations. In actual experiments, this is not going to be the case, and the timestream normalization will be determined as part of the calibration procedure. Any uncertainty of the procedure, as well as, that in the knowledge of the bandpassess will propagate to the final analysis products potentially further enhancing the residuals.
\section{Methodology} \label{sec:pipeline}

Before describing the pipeline, we summarize the instrumental assumptions 
adopted throughout this work. The HWP is the first optical element. Detectors are modeled as ideal linear polarizers (Eq.~\ref{eq:Mdet}). Beams are 
assumed perfectly collimated, with normal incidence on the HWP and no beam 
chromaticity.

\begin{figure}[ht!]
    \centering
    \includegraphics[width=\columnwidth]{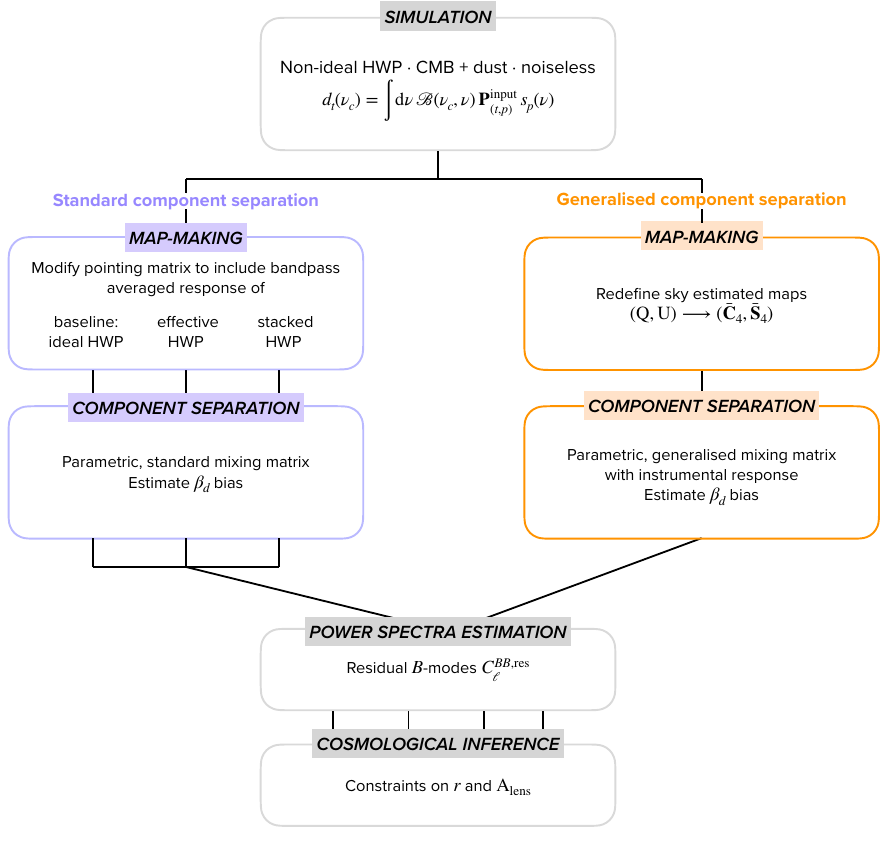}
    \caption{Overview of the two analysis pipelines studied in this work. Both 
    share a common simulation stage (top) and cosmological inference stage 
    (bottom). They differ in their treatment of HWP non-idealities: the standard 
    pipeline (left, purple) addresses instrumental effects by modifying the 
    pointing matrix at the map-making level, while the generalised pipeline 
    (right, teal) redefines the sky map estimates and incorporates the 
    instrumental response directly into the component-separation mixing matrix.}
    \label{fig:pipeline}
\end{figure}

We adopt a methodology closely related to the \texttt{xForecast} 
framework~\cite{stompor2016forecasting}. In this work, we explore two analysis pipelines, illustrated in 
Fig.~\ref{fig:pipeline}. Both of them share a common simulation stage: we 
begin by generating CMB and foreground simulations using 
\texttt{PySM}~\footnote{\href{https://pysm3.readthedocs.io/en/latest/}{\texttt{https://pysm3.readthedocs.io/en/latest/}}}, 
see Sec.~\ref{subsubsec:inputmaps}, and associate the effects of a telescope's 
pointing, see Sec.~\ref{subsec:obs_strat}, using a non-ideal HWP, 
$\mathbf{H}_{\rm{nonid}}$, introduced in Sec.~\ref{subsec:nonidealhwp}. More 
realistic simulations taking into account correlated noise and more complex Galactic 
foregrounds will be explored in future work.

The two pipelines then diverge at the map-making stage, as illustrated in 
Fig.~\ref{fig:pipeline}. In the \emph{standard pipeline} (left column), 
instrumental effects are addressed by modifying the pointing matrix, using either 
the effective or stack HWP model (Sec.~\ref{subsec:map-making}). The resulting maps 
are passed to a standard parametric component separation 
(Sec.~\ref{subsubsec:stdcompsep}), which yields foreground-cleaned CMB maps. 
Residuals are expected, as the data model does not fully account for all 
instrumental effects, leading to biased foreground SEDs and skewed cosmological 
parameters. In the \emph{generalised pipeline} (right column), the instrumental 
response is not assumed to be fully correctable at the map-making level. Instead, 
the map-making step redefines the meaning of the recovered sky maps by projecting 
the data onto the pure $4f$ harmonic basis, producing mixed-Stokes maps 
$\bar{\mathcal{C}}_4$ and $\bar{\mathcal{S}}_4$. These serve as inputs to a 
generalised component separation in which the frequency-dependent HWP mixing is 
incorporated directly into the mixing matrix 
(Sec.~\ref{subsubsection:gencs}). Both pipelines converge at the power spectrum 
estimation and cosmological inference stages, where we derive constraints on $r$ 
and $A_{\rm lens}$.

Since we are particularly interested in HWP-induced foreground residuals, we 
deliberately adopt \emph{noiseless} simulated datasets throughout this work.

\subsection{Simulation configuration} \label{subsec:simu}

\subsubsection{Input maps} \label{subsubsec:inputmaps}
For concreteness, we assume an experiment analogous to the SO-SAT \cite{yamada2024simons}, but focus on four frequency bands: MF1 (90 GHz), MF2 (150 GHz), UHF1 (220 GHz) and UHF2 (280 GHz). Each frequency band is modeled with a top-hat bandpass of 30\% width~\cite{Thornton_2016, Ward_2018}. The focal plane is fixed and consists of a square grid of $30 \times 30$ uniformly distributed detectors, covering a field of view corresponding to the central wafer. 

We simulate foregrounds considering only Galactic thermal dust emission, since our analysis is restricted to frequencies $\nu \geq 90$ GHz where polarized Galactic synchrotron emission is expected to be sub-dominant compared to dust~\cite{Krachmalnicoff_2018}. The latter is simulated using \texttt{PySM} \texttt{d0} model, i.e. following a modified blackbody spectrum, with fixed spectral parameters $\beta_d = 1.54$ and $T_d = 20$~K, both assumed to be spatially invariant. The resulting biases should be interpreted as lower bounds, or best-case scenarios, of what could happen in reality when all effects are included. CMB maps are produced with \textsc{Planck} 2018 cosmology \cite{Planck2020} for lensed scalar $B$-modes (A$_{\rm lens}=1$, $r$=0). Since \texttt{PySM} relies on the \texttt{healpy}~\footnote{\href{https://healpy.readthedocs.io/en/latest/}{\texttt{https://healpy.readthedocs.io/en/latest/}}} 
library following the \textsc{HEALPix} pixelization scheme~\cite{Gorski_2005}, we adopt a $\texttt{nside}=64$ resolution consistently throughout this study.

\subsubsection{Observing strategy}\label{subsec:obs_strat}

\begin{figure}[ht!]
    \centering
    \includegraphics[width=\columnwidth]{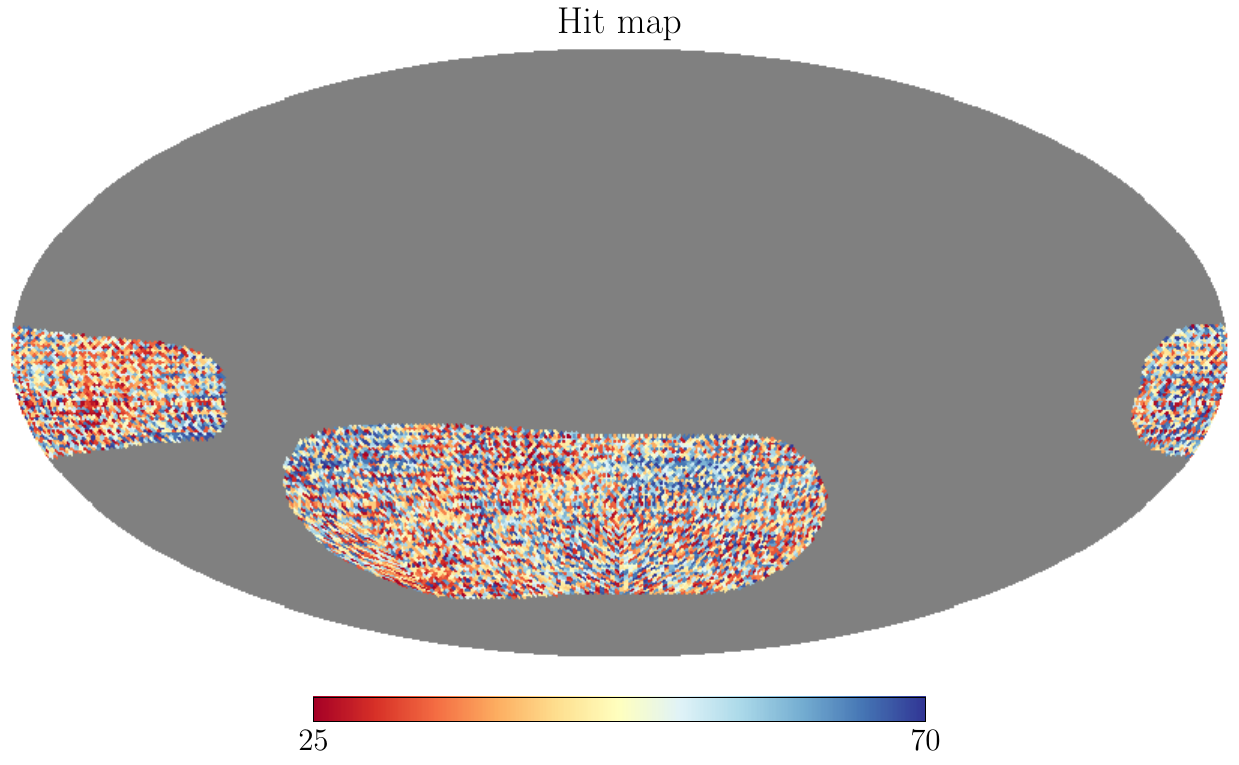}
    \caption{Sky hit map illustrating the distribution of detector observations with random sampling over a SO-like survey region.}
    \label{fig:hitmap}
\end{figure}

The following analysis focuses on large angular scales and adopts a simplified 
description of the observing strategy in order to  the impact of HWP 
non-idealities. As summarized in the instrumental assumptions above, we adopt 
ideal beam coupling, perfect collimation, and normal incidence on the HWP 
throughout. We adopt random uniform sampling rather than a realistic 
azimuth-constant-elevation scan. Since our simulations are noiseless, 
the scanning strategy does not affect the noise properties of the 
recovered maps, and the HWP-induced frequency-dependent effects we aim 
to isolate are independent of the specific scan pattern to leading 
order. A realistic scanning strategy will be considered in future work, 
where scan-synchronous systematics will also be assessed. The sky signal is nevertheless simulated over the full sky, and the 
corresponding mask is applied only to the output maps when forming the inputs 
to the component-separation analysis, Fig.~\ref{fig:hitmap}. We restrict the 
analysis to multipoles $\ell_{\max} < 200$, which encompass the angular scales 
carrying most of the information on primordial $B$-modes~\cite{Kamionkowski_2016, 
Planck2020}. This choice is also well 
matched to the resolution of our maps: at $N_{\rm side} = 64$ the \texttt{HEALPix} 
pixel scale is approximately $55$ arcmin, corresponding to a reliable multipole 
range up to $\ell \sim 2N_{\rm side} = 128$, while remaining computationally 
inexpensive.

We model the HWP as a multi-layer achromatic stack with parameters fixed to a SO-like design. In this model, the central layer is rotated by $\xi_{\rm MF} = 54^\circ$ for MF and $\xi_{\rm HF} = 56^\circ$ for UHF relative to the plate optical axis. All instrumental parameters defining this HWP model (including layer thicknesses and refractive indices) are taken from the specifications in~\cite{sakaguri2024antireflection}.

\begin{table}[h]
    \centering
    \resizebox{\columnwidth}{!}{%
    \begin{tabular}{lccc}
    \hline\hline
    \textbf{Material}               & \textbf{$ n $}                                      & $\theta_{\rm th}$ [mm] (MF)   & $\theta_{\rm th}$ [mm] (UHF)  \\ \hline
    Sapphire                       & 
    \begin{tabular}{c}
        $ n_o = 3.05 \pm 0.03 $ \\
        $ n_e = 3.38 \pm 0.03 $
    \end{tabular}
                                    & $ 3.75 \pm 0.01 $  & $ 1.60 \pm 0.01 $  \\
    Mullite                         & $ 2.52 \pm 0.02 $         & $ 0.212 \pm 0.01 $ & $ 0.097 \pm 0.01 $ \\
    Duroid                          & $ 1.41 \pm 0.01 $         & $ 0.394 \pm 0.01 $ & $ 0.183 \pm 0.01 $ \\
    Epo-Tek (glue)                  & $ 1.7 $                    & $ 0.04 $           & $ 0.04 $           \\ 
    \hline\hline
    \end{tabular}%
    }
    \caption{Refractive indices and effective thicknesses of the HWP stack from \cite{sakaguri2024antireflection}.}
    \label{tab:instparams}
\end{table}

\subsection{Map-making procedure}
\label{subsec:map-making}
The data model particularly suitable for the map-making procedure, e.g., ~\cite{poletti2017}, is linear in the sky signal amplitude, where the signal part of each measurement can be modeled as 

\begin{equation}
    \mathbf{d}_t = \mathbf{P}_{(t,p)} \, \mathbf{s}_p
    \label{eq:datamodel0}
\end{equation}
where $\mathbf{d}_t$ is the time-ordered data (TOD), $\mathbf{s}_p$ is the sky signal in pixel $p$. 
In the case at hand the standard map-making pointing matrix, $\mathbf{P}_{(t,p)}$, has to be generalized in order to capture not only the scanning details but also the full instrumental 
response, including polarization modulation by the HWP and other optical elements.
As these depend on the frequency the actual data model is given as in Eq.~\ref{eq:datamodel1}, which, while also linear, is not directly in the form of Eq.~\ref{eq:datamodel0}. To facilitate the map-making procedure we have to recast this latter equation into the form of the former. We can do that either by modifying the generalized pointing operator, so that the corresponding sky map better approximates the actual sky, or, alternately, we can redefine the meaning of the sky maps estimated by the procedure. We pursue both these options in this work. We note that the former approach can only be approximate. Its main advantage is that it requires only minimal modifications to the standard map-making pointing matrix, leaving the rest of the analysis pipeline unchanged.\\ 
\subsubsection{An effective HWP model}
We consider two cases of approximate pointing matrices.
In the first case, we assume that all the effects due to the inclusion of the Mueller matrix in the pointing operator can be modeled via a simple phase shift added to the pointing matrix of the ideal HWP case, Eq.~\ref{eq:idtod}, i.e., 
\begin{equation}
    \begin{split}
    \mathbf{d}^{\mathrm{eff}}_{t}(\nu_c) = \varepsilon(\nu_c)\, \Big[ 
    & \mathrm{I} _t + \mathrm{Q}_t \cos\big(4\varphi_t + 2\alpha_t +\phi(\nu_c)\big) \\
    & + \mathrm{U}_t \sin\big(4\varphi_t + 2\alpha_t + \phi(\nu_c)\big) \Big],
    \end{split}
    \label{eq:deff}
\end{equation}
where $\mathrm{I}_t$, $\mathrm{Q}_t$, and $\mathrm{U}_t$, describe the sky signal observed at time $t$, and  we have introduced two parameters, a phase shift $\phi(\nu_c)$ and an amplitude factor $\varepsilon(\nu_c)$, both defined for each frequency band centered at $\nu_c$. These aim at mimicking the effects of the HWP and we determine them both by fitting directly the expression in Eq.~\ref{eq:deff} to the non-ideal time-ordered data in each frequency band, i.e.,
\begin{equation}
    \{\varepsilon(\nu_c), \phi(\nu_c)\} = 
    \arg\min_{\varepsilon,\phi}\, 
    \left\| \mathbf{d}^{\mathrm{nonid}}_{t}(\nu_c) - \mathbf{d}^{\mathrm{eff}}_{t}(\nu_c) \right\| ,
    \label{eq:fit}
\end{equation}
where the non-ideal band-integrated signal is defined as
\begin{equation}
    \mathbf{d}^{\mathrm{nonid}}_{t}(\nu_c) =
    \int d\nu \, \boldsymbol{\mathcal{B}}(\nu_c, \nu)\,
    \mathbf{M}^{\rm nonid}_t(\nu)\, \mathbf{s}(\nu) ,
    \label{eq:dsnonid}
\end{equation}
with $\boldsymbol{\mathcal{B}}(\nu_c,\nu)$ the bandpass centered at $\nu_c$, 
$\mathbf{M}^{\rm nonid}_t(\nu)$ the Mueller matrix of the non-ideal HWP, and $\mathbf{s}(\nu)=(\mathrm{I},\mathrm{Q},\mathrm{U})=(0,1,0)$ the input Stokes vector, chosen to isolate the modulated polarization component used for the fit. This procedure mimics an actual calibration procedure as could be performed in the field employing an artificial source  with known output brightness and polarization. Examples of such procedures include the use of a wire-grid polarizer or other polarized calibration sources~\cite{nakata2025, Coppi_2025}.
\\
We refer to this model as an effective HWP data model.

\subsubsection{A stacked HWP model.} \label{subsubsec:mapmakingstack}
As the second case of the modified pointing we consider the model also based on the three-layer stack HWP but this time we average the Mueller matrix elements over the bandpasses and use those to define the generalized pointing matrix for each bandpass.
This effectively corresponds to replacing the stack Mueller matrix, $\mathbf{H}_{\mathrm{stack}}$, in the expression for the Mueller matrix of the full optics chain in Eq.~\ref{eq:Mstack} with their bandpass integrated version, $\mathbf{\bar H}_{\mathrm{stack}}$, given by,
\begin{align}
\mathbf{\bar H}_{\mathrm{stack}}(\nu_c, \gamma)
&\equiv
\int d\nu\,
\boldsymbol{\mathcal{B}}(\nu_c, \nu)\,
\mathbf{H}_{\mathrm{stack}}(\nu, \gamma)\,
\mathcal{F}(\nu, \nu_c)
\label{eq:stackbpav}
\end{align}
where $\gamma$ as before stands here for the physical parameters of the HWP, i.e., thicknesses of each of the layers in the case at hand, and we assume that the incident light falls onto the HWP always perpendicularly. The additional frequency-dependent factor, $\mathcal{F}(\nu, \nu_c)$, which can be frequency-band specific, is optional but could improve the performance of the approximation. Indeed, if it were possible to capture the frequency dependence of the actual sky signal, with such a single, a priori known function, this would allow for an accurate estimation of the sky maps for each frequency band averaged by the bandpass modified by this dependence. These maps would contain all the information about the sky which could be then passed to the component separation method as it is the case in the standard pipeline. However, the actual scaling of the sky is not only more complex, due to its multi-component nature, but it is also pixel-dependent and a priori unknown. Nevertheless, some approximate guess can already be helpful, and we as an example, consider cases where we adopt as $\mathcal{F}$ the scaling of what we guess is the dominant component at each of the frequency bands, see Sec.\ref{sec:appdust}

\begin{figure}[ht!]
    \centering
    \includegraphics[width=\columnwidth]{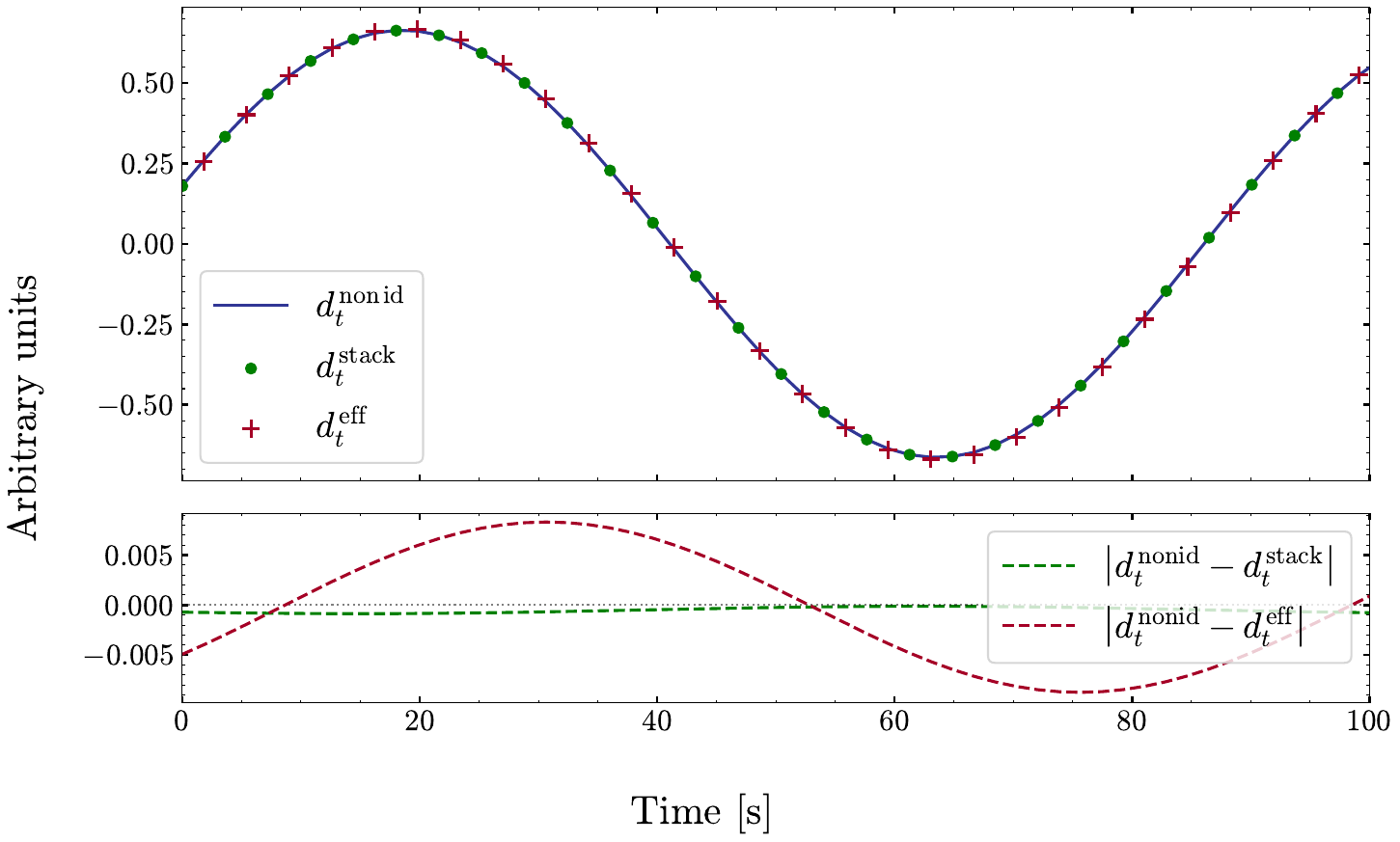}
    \caption{Bandpass-integrated time-ordered data (TOD) in the 220~GHz band for a fixed, linearly polarized input signal with Stokes parameters $\mathrm{I}=0$, $\mathrm{Q}=1$, $\mathrm{U}=0$ assuming $\alpha_t=0$ and the HWP rotating uniformly with $f_{\mathrm{HWP}} =2$Hz. The TOD is shown for the three HWP models, $\mathbf{M}_{\rm nonid}$, $\mathbf{M}_{\rm stack}$, and $\mathbf{M}_{\rm eff}$, together with the corresponding residuals with respect to the non-ideal model.}
    \label{fig:todconst}
\end{figure}

While such a model induces on its own a phase shift of the polarized light modulation, this is not sufficient to account fully for the phase shift we see in the simulated data. Therefore, for each frequency band, we introduce two additional phenomenological parameters, the additional phase shift, $\phi^\prime(\nu_c)$ and the gain factor, $\varepsilon^\prime(\nu_c)$. These directly modify the arguments of the cosines and sines describing the HWP signal modulation as in the effective model case, Eq.~\ref{eq:deff}, and they are  fitted for as part of the calibration procedure as discussed earlier. An example of such a fit-corrected time-stream is shown in Fig.~\ref{fig:todconst}. 
\\
Algebraically, these extra parameters are defined as follows. We first observe, see e.g.,~\cite{verges2021framework}, that $(0,0)$ element of the HWP stack Mueller matrix in Eq.~\ref{eq:Mopticchainstack}, ${\mathbf M}^{\mathrm{stack}}_{0i}$, is time-independent while the polarized ones, i.e., $(0,1)$ and $(0,2)$ can be represented as a series of terms with either $\cos( k\varphi_t + 2\alpha_t)$ or $\sin (k\varphi_t+\alpha_t)$, where $k=0$ or $k=4$, i.e.,
\begin{align}
    \mathbf{M}^{\rm stack}_{0i} &= \sum\limits_{k=0,4} \mathbf{C}^{\mathrm{stack}}_{0i,k}(\nu, \gamma)\,
        \cos\!\big(k\varphi_t + 2 \alpha_t\big) \nonumber \\
    &\quad + \sum\limits_{k=0,4} \mathbf{S}^{\mathrm{stack}}_{0i,k}(\nu, \gamma)\,
        \sin\!\big(k\varphi_t + 2 \alpha_t\big).
    \label{eq:Mharmonics}
\end{align}

\noindent In the following we are only interested in the polarized components modulated at 4 times the HWP frequency, and we thus retain only the $k=4$ terms. We can use then the above equation to write down the data model for the three layer HWP replacing the frequency-dependent coefficients with their bandpass averaged values, marked with a bar. This gives us,
\begin{align}
       \mathbf{d}^{\mathrm{stack}}_{t}(\nu_c) \; \simeq \;   \Big[ \mathrm{I}_t & + \mathrm{Q}_t \; \mathbf{\bar{C}}^{\mathrm{stack}}_{01,4}(\nu_c, \gamma) \cos (4\varphi_t+2\alpha_t)\nonumber\\
        & + \mathrm{Q}_t\; \mathbf{\bar{S}}^{\mathrm{stack}}_{01,4}(\nu_c, \gamma)\sin (4\varphi_t+2\alpha_t)\nonumber\\
   & + \mathrm{U}_t \;\mathbf{\bar{C}}^{\mathrm{stack}}_{02,4}(\nu_c, \gamma) \cos (4\varphi_t+2\alpha_t) \nonumber\\
   & + \mathrm{U}_t \; \mathbf{\bar{S}}^{\mathrm{stack}}_{02,4}(\nu_c, \gamma) \sin (4\varphi_t+2\alpha_t)\nonumber
    \end{align}

This can be rewritten as
\begin{align}
\mathbf{d}^{\mathrm{stack}}_{t}(\nu_c) & = \mathrm{I}_t  
+ \mathrm{Q}_t\, \eta_Q(\nu_c)\,
\cos (4\varphi_t+2\alpha_t + \psi_Q(\nu_c)) \nonumber \\
&\quad + \mathrm{U}_t \,\eta_U(\nu_c)\,
\sin (4\varphi_t+2\alpha_t + \psi_U(\nu_c))
\label{eq:fitstack}
\end{align}
where extra functions $\eta_{Q/U}(\nu_c)$ and $\psi_{Q/U}(\nu_c)$ can all be related to the HWP physical parameters via the bandpass averaged versions of the coefficients, $\mathbf{C}^{\mathrm{stack}}_{0i,k}(\nu, \gamma)$ and $\mathbf{S}^{\mathrm{stack}}_{0i,k}(\nu, \gamma)$. The detailed expressions are developed in App.~\ref{app:amp_phase}. This model can be further extended by adding the phenomenological parameters introduced earlier to read,
\begin{align}
        \mathbf{d}^{\mathrm{stack}}_{t}(\nu_c)  \; = &\; \varepsilon'(\nu_c) \, \times \label{eq:dstack}\\
        \times\,\Big[ \mathrm{I}_t 
         + \mathrm{Q}_t \, &\eta_Q(\nu_c) \; 
        \cos (4\varphi_t+2\alpha_t + \psi_Q(\nu_c) \, + \, \phi'(\nu_c))  \nonumber \\
       + \mathrm{U}_t \, &\eta_U(\nu_c) \; \sin (4\varphi_t+2\alpha_t + \psi_U(\nu_c) + \phi'(\nu_c))\Big].\nonumber
\end{align}
This is the expression which should be directly contrasted with the expression in Eq.~\ref{eq:deff} defining the effective model. We emphasize that in Eq.~\ref{eq:dstack} only the parameters $\varepsilon'(\nu_c)$ and $\phi'(\nu_c)$ are fitted for while all the others are inferred from the HWP model and its physical parameters. A potentially interesting alternative could be to fit for all these 5 parameters, i.e., two arbitrary phases shifts and three amplitudes, independently. We leave this here for future exploration.

\begin{figure}[ht!]
    \centering
    \includegraphics[width=\columnwidth]{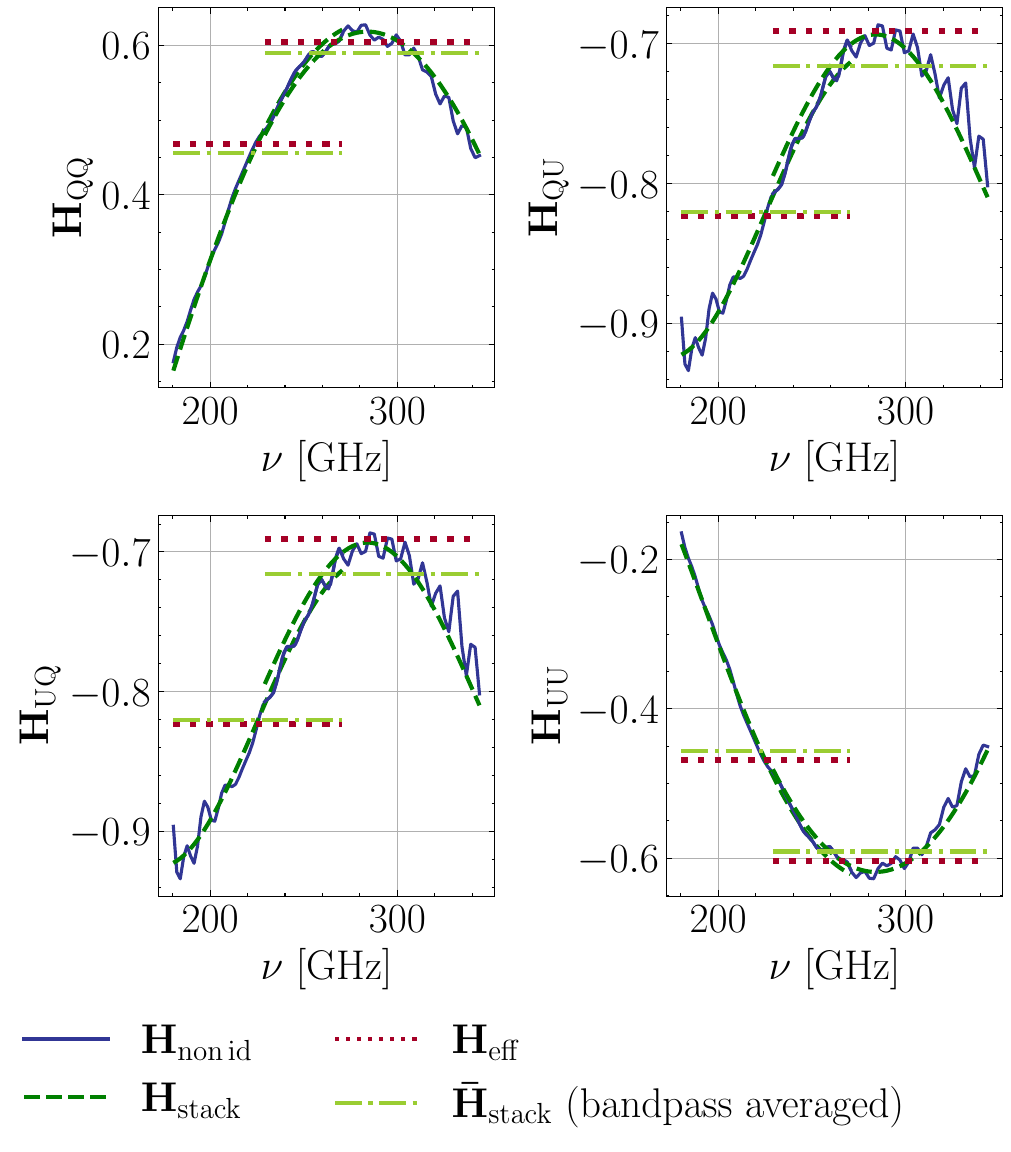}
    \caption{Frequency dependence in the UHF band of the central $2\times2$ $(\mathrm{Q},\mathrm{U})$ block of the HWP Mueller matrix for two HWP models considered in this work: the full non-ideal model $\mathbf{H}_{\rm nonid}$ and the stacked model $\mathbf{H}_{\rm stack}$. The horizontal lines show: the effective model, $\mathbf{H}_{\rm eff}$, red dots, and bandpass-averaged stack model, $\mathbf{\bar{H}}_{\rm stack}$, light green dashes, as used in the map-making pointing model.}
    \label{fig:Miifreq}
\end{figure}

Fig.~\ref{fig:Miifreq} shows the frequency dependence of the central $2\times2$ $(\mathrm{Q},\mathrm{U})$ block of the Mueller matrix for the three HWP models across the UHF band. The HWP Mueller matrix elements shown here correspond to the $\mathbf{H}_{\rm X}$ block entering the time dependent Mueller matrix of the full optical chain, $\mathbf{M}^{\rm X}_t$ (Eq.~\ref{eq:Mopticchain}). In the time-domain simulations we use the full frequency-dependent non-ideal HWP model (\textit{solid navy line}). In the map-making step, however, we employ bandpass-averaged approximations of the models as described above, shown in horizontal lines. The bandpass-averaged stacked model, shown as the\textit{ light-green dashed line}, correctly captures some of the difference in amplitude between the different Mueller matrix elements (Eq.~\ref{eq:dstack}), resulting in a more accurate representation of the underlying matrix. It allows to suppress the time-domain residuals roughly an order of magnitude better than the effective model, as illustrated in Fig.~\ref{fig:todconst}.

While in both cases the time domain residuals are quite low it remains to be seen whether it is sufficient to enable a high-fidelity recovery of the CMB signal. We study this in the reminder of this paper.
We also note that whenever any of these two approximations works sufficiently well, the estimated sky signal should be close to the bandpass smoothed sky signal. This expectation however does not have to be the case in general and need to be verified case-by-case. 

\subsubsection{The generalized sky map model.}
The second option is more elaborated as it combines effectively map-making and component separation stages by redefining the sky amplitudes so they incorporate all frequency-dependent factors, averaged over the frequency bands and the sky signals. These new generalized sky signal will now incorporate linear combinations of different Stokes parameters of different sky components. This approach was first discussed in \cite{verges2021framework}. Its advantage is that it can be as accurate as the Mueller matrix models used. The downside is that it does not directly produce physically meaningful maps of the sky and that it needs to be complemented by a generalized, and potentially more complex component separation procedure capable of dealing with more complex input sky signals~\cite{verges2021framework}. We describe that latter procedure later on and focus here only on the map-making. For more details, we refer the reader to ~\cite{verges2021framework}.\\
As before we start with the detector model for the monochromatic light, which can be written as,
\begin{align}
    \mathbf{d}^{\mathrm{X}}_{t}(\nu) &= {\mathbf M}^{\rm X}_{00}(\nu, \gamma)\, \mathrm{I}_t(\nu) \nonumber \\
    &\quad + {\mathbf M}^{\rm X}_{01}(t, \nu, \gamma)\, \mathrm{Q}_t(\nu)\nonumber \\
    &\quad + {\mathbf M}^{\rm X}_{02}(t, \nu, \gamma)\, \mathrm{U}_t(\nu),
    \label{eq:dQUmix}
\end{align}
where $\gamma$ are the instrumental parameters introduced in Sec.\ref{sec:instmod} and the subscript $\rm X$ denotes the adapted polychromatic HWP model, which hereafter is either the stack or the non-ideal model. We assume normal incidence throughout, consistent with the instrumental assumptions stated at the beginning of Sec.~\ref{sec:pipeline}.
The Stokes parameters, $\mathrm{I}_t(\nu)$, $\mathrm{Q}_t(\nu)$, and $\mathrm{U}_t(\nu)$ denote the sky signal at frequency $\nu$ as measured at time $t$. We then observe that for any HWP model whose properties do not inherently depend on the HWP orientation relative to either the sky or the instrument frame, all the time dependence of the Mueller matrix coefficients,
${\mathbf M}^{\rm X}_{0i}$, is only due to the rotation matrices, e.g., Eq.~\ref{eq:Mopticchain}. Consequently, the $(0,0)$ element is a constant while the polarized ones can be represented as a series of terms with either $\cos (k\varphi_t + 2\alpha_t)$ or $\sin (k\varphi_t+\alpha_t)$, where $k=0$ or $=4$ as in Eq.~\ref{eq:Mharmonics}.

Upon inserting these relations into the data model, Eq.~\ref{eq:dQUmix}, and integrating the signal over the frequency band pass for each band, we obtain, 
\begin{align}
    \mathbf{d}^{\mathrm{X}}_{t}(\nu_c) & \equiv
    \int d\nu \, \boldsymbol{\mathcal{B}}(\nu_c,\nu)\, \mathbf{d}^{\mathrm{X}}_{t}(\nu) \nonumber \\
    & \quad = \; 
    \int d\nu \, \boldsymbol{\mathcal{B}}(\nu_c,\nu)\, \mathrm{I}_t(\nu) \nonumber \\
    &\quad + 
    \sum_{k=0,4}\,  \boldsymbol{\mathcal{C}}_{k, t}(\nu_c,\gamma) \, \cos(k\varphi_t+2\alpha_t)\,
    \nonumber
    \\
    &\quad + 
    \sum_{k=0,4}\,  \boldsymbol{\mathcal{S}}_{k, t}(\nu_c,\gamma)\,\sin(k\varphi_t+2\alpha_t),
    \label{eq:stack_model}
\end{align}
where,
\begin{align}
    \boldsymbol{\mathcal{C}}_{k, t}(\nu_c,\gamma) \; \equiv \; \Big[\int & d\nu \, \boldsymbol{\mathcal{B}}(\nu_c,\nu)\,
    {\mathbf C}^{\rm X}_{01,k}(\nu, \gamma)\, \mathrm{Q}_t(\nu) 
    \label{eq:CsampleDef}
    \\ & + \boldsymbol{\mathcal{B}}(\nu_c,\nu)\,
    {\mathbf C}^{\rm X}_{02,k}(\nu, \gamma)\, \mathrm{U}_t(\nu)\Big]\nonumber\\
    \boldsymbol{\mathcal{S}}_{k, t}(\nu_c,\gamma)  \; \equiv \; \Big[\int & d\nu \, \boldsymbol{\mathcal{B}}(\nu_c,\nu)\,
    {\mathbf S}^{\rm X}_{01,k}(\nu, \gamma)\, \mathrm{Q}_t(\nu)
    \label{eq:SsampleDef}
    \\
    & + \boldsymbol{\mathcal{B}}(\nu_c,\nu)\,
    {\mathbf S}^{\rm X}_{02,k}(\nu, \gamma)\, \mathrm{U}_t(\nu)\Big]. \nonumber
\end{align}
As a result, the polarization signal is no longer encoded purely in the $\cos(4\varphi_t+2\alpha_t)$ and $\sin(4\varphi_t+2\alpha_t)$ modulations uniquely associated with $\mathrm{Q}$ and $\mathrm{U}$ in the ideal case (Eq.~\ref{eq:idtod}). Moreover, it cannot be modeled as a simple, sky-signal-independent phase shift, as would be the case for monochromatic light or as assumed in the stack approximation introduced earlier (Eq.~\ref{eq:dstack}). However, assuming the model defined by Eq.~\ref{eq:stack_model}, and interpreting the coefficients 
$\boldsymbol{\mathcal{C}}_{k, t}(\gamma)$ and $\boldsymbol{\mathcal{S}}_{k, t}(\gamma)$ as new, redefined sky signals, we can use the map-making procedure to recover them with the pointing matrix
uniquely defined by the cosines and sines functions.
\\
In the following we will further simplify Eq.~\ref{eq:stack_model}, by neglecting the sky-only modulated components, i.e., total intensity and $k=0$ terms. This is justified by our focus on the polarization signals. We will also neglect possible presence of any additive effects such as HWP-synchronous signals. These should and could be mitigated during the map-making and will be included in a follow-up analysis~\cite{Sohn2026}.  We note that this simplification is also implicitly made in the standard  pipeline approach of Sec.~\ref{subsec:map-making}, where the map-making procedure retains only the 4f harmonic and the intensity term is not used in polarization recovery. The comparison between standard and generalized approaches is therefore not affected by this difference in assumption; both operate on the polarized $4f$-modulated signal. The mixed-Stokes sky signals recovered from the map-making are then related to the standard Stokes parameters as we discuss in Sec.~\ref{subsubsection:gencs}.

We emphasize that as this approach aims at providing a complete model of the HWP effects, it does not require any ad-hoc phenomenological parameters, as the approximate models introduced earlier. In practice, such parameters may still be required, for instance, in order to account for the zero angle knowledge uncertainty or due to other elements of the optical chains, and could be readily introduce in the model, if needed, potentially improving its performance. They are however not included in the following discussion.

\subsubsection{Map estimation.}
In all these cases the corresponding sky signal is subsequently reconstructed from the time-ordered data using a generalized least-squares (GLS) map-making formalism. The corresponding estimator is,
\begin{equation}
    \hat{\mathbf{s}}^{\rm X}_p
    =
    \left(
    \mathbf{P}^{\mathrm{model}\,T}_{\rm X,(t,p)}
    \mathbf{N}^{-1}
    \mathbf{P}^{\mathrm{model}}_{\rm X,(t,p)}
    \right)^{-1}
    \mathbf{P}^{\mathrm{model}\,T}_{\rm X,(t,p)}
    \mathbf{N}^{-1}
    \, \mathbf{d}_t,
    \label{eq:mapmakingsolbis}
\end{equation}
 here $\mathbf{N}$ is the noise covariance matrix of the time-ordered data. Throughout this work, we assume uncorrelated noise with unit variance, $\mathbf{N}=\mathbf{I}$, such that Eq.~\eqref{eq:mapmakingsolbis} reduces to the standard normal-equation solution. The pointing matrix, $\mathbf{P}^{\mathrm{model}}_{\rm X}$, depends on which of the three options, the effective model, $\mathrm{X}=\mathrm{eff}$, the stacked model, $\mathrm{X}=\mathrm{stack}$, or the generalized solution, $\mathrm{X}=\mathrm{gen}$, is considered. In addition, as a reference, we will consider the pointing matrix corresponding to a perfect, monochromatic HWP, $\mathrm{X}=\mathrm{ideal}$, as given in Eq.~\ref{eq:idtod}.
 
Once the maps, $\hat{\mathbf{s}}^{\rm X}$, are reconstructed, we evaluate, for each of the schemes studied here,
how accurately they reproduce the input component maps and  quantify the 
systematic biases induced by an imperfect modeling of the HWP in the map-making procedure.

\subsection{Parametric component separation} \label{subsec:compsep}

\subsubsection{Standard approach}
\label{subsubsec:stdcompsep}
    
We adopt a parametric component-separation model in which the multi-frequency polarized sky signal is described as a superposition of astrophysical components. Under this assumption, the polarization maps reconstructed at different observing frequencies can be written, for each pixel $p$, as

\begin{equation}
    \hat{\mathbf{s}}_p(\nu) = \mathbf{A}(\nu,\nu_0)\, \mathbf{s}^{\rm comp}_p(\nu_0),
    \label{eq:dAs}
\end{equation}

\noindent where $\hat{\mathbf{s}}_p$ is the vector collecting the reconstructed Stokes polarization maps at all frequency bands,

\begin{equation}
    \hat{\mathbf{s}}_p =
    (\mathrm{Q}_p(\nu_1), \mathrm{U}_p(\nu_1), \ldots, \mathrm{Q}_p(\nu_f), \mathrm{U}_p(\nu_f))^{\rm T},
\end{equation}

\noindent and $\mathbf{s}^{\rm comp}_p(\nu_0)$ contains the Stokes parameters of the astrophysical components evaluated at a reference frequency $\nu_0$,

\begin{multline}
    \mathbf{s}^{\rm comp}_p(\nu_0) = \big(\mathrm{Q}_p^{\rm CMB}(\nu_0),\, 
    \mathrm{U}_p^{\rm CMB}(\nu_0), \\
    \mathrm{Q}_p^{\rm dust}(\nu_0),\, 
    \mathrm{U}_p^{\rm dust}(\nu_0)\big)^{\rm T}.
\end{multline}

\noindent The mixing matrix $\mathbf{A}(\nu,\nu_0)$ describes the frequency scaling of each component. Assuming identical spectral scaling for the $\mathrm{Q}$ and $\mathrm{U}$ Stokes parameters, it can be written as

\begin{align}
\mathbf{A}(\nu,\nu_0) =
\begin{bmatrix}
a^{\rm CMB}(\nu_1,\nu_0)\mathbf{I}_2 & a^{\rm dust}(\nu_1,\nu_0)\mathbf{I}_2 \\
\vdots & \vdots \\
a^{\rm CMB}(\nu_f,\nu_0)\mathbf{I}_2 & a^{\rm dust}(\nu_f,\nu_0)\mathbf{I}_2
\end{bmatrix},
\label{eq:compsep}
\end{align}

\noindent where $\mathbf{I}_2$ is the $2\times2$ identity matrix acting on the $(\mathrm{Q},\mathrm{U})$ subspace and $a^{\rm CMB}$ and $a^{\rm dust}$ describe the frequency scaling of the CMB and dust components.

In the case of a parametric component 
separation, the coefficients $a^{\rm{comp}}(\nu_i, \nu_0)$ of 
the mixing matrix are assumed to be paramatrized by a 
few unknown parameters, $\beta_i$ called spectral parameters. By adopting this data model, Eq.~\ref{eq:dAs}, the component separation problem can be reformulated within a maximum likelihood framework to estimate the parameters 
$\beta_i$ that characterize each component spectral and spatial behaviour. Following ~\cite{stompor2009maximum} the spectral likelihood of the data reads as, 

\begin{multline}
-2\log\mathcal{L}_{\rm spec}(\beta) = \rm{const} \\ 
- \left(\mathbf{A}^T \mathbf{N}^{-1} \mathbf{\hat{s}_p}\right)^{T}  \left(\mathbf{A}^T \mathbf{N}^{-1} \mathbf{A}\right)^{-1} 
\left(\mathbf{A}^T \mathbf{N}^{-1} \mathbf{\hat{s}_p}\right)
    \label{eq:lspec}
\end{multline}

\noindent In this analysis we assume a diagonal noise covariance 
$\mathbf{N}_{ii} = 1\,\mu{\rm K}^2/{\rm pixel}$, corresponding to 
approximately $60\,\mu{\rm K}$-arcmin, which is comparable to the 
expected noise level of the SO UHF channels after $\lesssim 1$ year of 
integration. We stress that the simulated maps are noiseless; this 
fiducial covariance is introduced solely to obtain finite statistical 
uncertainties on the recovered spectral parameters.

\noindent Eq.~\ref{eq:lspec} reaches its maximum for $\tilde{\beta}$ and the amplitude of each sky component is given by the generalised least squares solution

\begin{equation}
    \tilde{\mathbf{s}}^{\rm{comp}}_p = \left(\tilde{\mathbf{A}}^T 
    \mathbf{N}^{-1} \tilde{\mathbf{A}}\right)^{-1} \tilde{\mathbf{A}}^T  
    \mathbf{N}^{-1} \hat{\mathbf{s}}_p \label{eq:gls}
\end{equation}

\noindent where $\tilde{\mathbf{A}} = \mathbf{A}(\tilde{\beta})$. In practice, each frequency map corresponds to a finite instrumental bandpass. The elements of the mixing matrix therefore represent bandpass-integrated spectral scalings rather than monochromatic values,

\begin{equation}
    a^{\rm comp}(\nu_c, \nu_0)  
    \equiv
    \int d\nu\;
    \boldsymbol{\mathcal{B}}(\nu_c,\nu)\,
    f^{\rm comp}_{\beta}(\nu, \nu_0),
\end{equation}

\noindent where $\boldsymbol{\mathcal{B}}(\nu, \nu_i)$ is the bandpass of channel centered at $\nu_c$, and $f^{\rm comp}_{\beta}(\nu, \nu_0)$ denotes the component spectral law evaluated at frequency $\nu$ and parameterized by the spectral parameters $\beta$.

\subsubsection{Generalised component separation}
\label{subsubsection:gencs}

The map-making procedure in the generalized case, will produce pixelized maps of the generalized sky signals via Eq.~\ref{eq:mapmakingsolbis}. There are two maps for each frequency band, $\nu_c$, denoted $\boldsymbol{\mathcal{C}}^{\rm X}_{4 p}(\gamma, \nu_c)$ and $\boldsymbol{\mathcal{S}}^{\rm X}_{4 p}(\gamma, \nu_c)$, each of which is a bandpass integrated, linear combination of the standard Stokes parameters $\mathrm{Q}$ and $\mathrm{U}$ characterizing the sky signal. They are defined in Eqs.~\ref{eq:CsampleDef} and~\ref{eq:SsampleDef}. 

Using Eq.~\ref{eq:compsep} we can rewrite the expressions for the mixed Stokes signals as,

\begin{align}
\begin{pmatrix}
\boldsymbol{\mathcal{C}}^{\rm X}_{4p}(\nu_c, \gamma)\\[6pt]
\boldsymbol{\mathcal{S}}^{\rm X}_{4p}(\nu_c, \gamma)
\end{pmatrix}
& = \int d\nu \, \boldsymbol{\mathcal{B}}(\nu_c, \nu) 
\begin{pmatrix}
\mathbf{C}^X_{01,4} & \mathbf{C}^X_{02,4}\\[4pt]
\mathbf{S}^X_{01,4} & \mathbf{S}^X_{02,4}
\end{pmatrix} \nonumber \\
& \hspace{4em} \times \; \mathbf{A}_p(\nu; \nu_0) \;
\mathbf{s}^{\mathrm{comp}}(\nu_0) \nonumber\\[6pt]
& \equiv \;
\boldsymbol{\mathcal{M}}^{\rm X}(\gamma, \nu_c) \,
\mathbf{s}^{\mathrm{comp}}_p(\nu_0),
\end{align}
\noindent where the coefficients $\mathbf{C}^X_{0i,4}$ and 
$\mathbf{S}^X_{0i,4}$ depend on $(\nu, \gamma)$ and are integrated 
over the bandpass together with $\boldsymbol{\mathcal{B}}(\nu_c, \nu)$. This equation has a form appropriate for the standard component separation procedure with a new mixing matrix given by $\boldsymbol{\mathcal{M}}^{\rm X}(\gamma, \nu_c)$.

Our entire procedure is now as follows.
We first explicitly recover the mixed-
Stokes maps $\boldsymbol{\mathcal{C}}_4$ and $\boldsymbol{\mathcal{S}}_4$ at the 
map-making stage and use them as input to the component-
separation analysis described in 
Sec.~\ref{subsec:appgencs}, Fig.\ref{fig:pipeline}. This approach generalizes 
the standard component-separation framework by treating 
instrumental polarization mixing and astrophysical 
emission jointly within a unified data 
model. 

% \begin{figure}[ht!]
%     \centering
%     \includegraphics[width=\columnwidth]{figures/pipeline_bis.pdf}
%     \caption{Methodology for the generalised component separation. Simulated data are generated including non-ideal HWP modulation, and the $4f$ harmonic bandpass averaged maps $(\boldsymbol{\mathcal{C}}_4,\boldsymbol{\mathcal{S}}_4)$ are used as inputs to the component-separation pipeline. Instrumental effects are accounted for by incorporating the HWP response directly into the mixing matrix.}
%     \label{fig:pipeline_bis}
% \end{figure}

\subsection{Power spectra and cosmological likelihood}

To quantify the level of residual contamination in the reconstructed maps, we define, for each HWP model $\mathrm{X}$, the residual sky signal as the difference between the component-separated reconstructed map and the corresponding input CMB signal,
\begin{equation}
    s^{\mathrm{X}}_{\rm res}
    \equiv
    \hat{s}^{\mathrm{X}} - s^{\rm CMB},
    \label{eq:resmap}
\end{equation}
where the explicit pixel index $p$ has been omitted for clarity. We then compute the associated residual $B$-modes power spectrum for each HWP model X using the \texttt{NaMaster}~\footnote{\href{https://namaster.readthedocs.io/en/latest/}{\texttt{https://namaster.readthedocs.io/en/latest/}}} library,
\begin{equation}
    C_{\ell,\,\mathrm{res}}^{{BB},\mathrm{X}}
    \equiv
    C_\ell^{BB}\!\left(s^{\mathrm{X}}_{\rm res}\right).
    \label{eq:rescl}
\end{equation}

Building on the residual $B$-modes spectra discussed above, we propagate these residuals into cosmological parameter estimation by sampling the posterior distribution of the parameters $\boldsymbol{\theta}\equiv\{r, A_{\rm lens}\}$, 
\begin{equation}
    P(\boldsymbol{\theta} \mid \hat{\mathbf{C}}^{BB})
    \propto
    \mathcal{L}(\hat{\mathbf{C}}^{BB}\mid \boldsymbol{\theta})\,\pi(\boldsymbol{\theta}),
\end{equation}
where $\pi(\boldsymbol{\theta})$ denotes the adopted priors and
$\hat{\mathbf{C}}^{BB}$ is the data vector formed by the measured binned
$B$-modes power spectrum,
\begin{equation}
    \hat{\mathbf{C}}^{BB}
    \equiv
    \left(
        \hat{C}^{BB}_{b_1},
        \hat{C}^{BB}_{b_2},
        \ldots,
        \hat{C}^{BB}_{b_{N_{\rm bin}}}
    \right)^{T}.
\end{equation}

\noindent The corresponding model prediction is written as
\begin{equation}
    \mathbf{C}^{BB}(\boldsymbol{\theta})
    =
    \mathbf{C}^{BB}_{\rm th}(r, A_{\rm lens})
    +
    \mathbf{C}^{BB}_{\rm res},
\end{equation}
\noindent where $\mathbf{C}^{BB}_{\rm res}$ is the residual $B$-modes spectrum estimated
from the component-separation pipeline for the corresponding HWP model.
The theoretical $B$-modes spectrum is modeled as
\begin{equation}
    C_{\ell,\,\rm th}^{BB}(r, A_{\rm lens})
    =
    r\,C_{\ell,\,\rm prim}^{BB}
    +
    A_{\rm lens}\,C_{\ell,\,\rm lens}^{BB},
    \label{eq:BBth}
\end{equation}
\noindent with $C_{\ell,\,\rm prim}^{BB}$ and $C_{\ell,\,\rm lens}^{BB}$ denoting the
primordial and lensing $B$-mode templates, respectively. We model the likelihood of the binned power spectra as
\begin{multline}
    -2\ln \mathcal{L} =
    \sum_{b}
    (2\ell_{b}+1)\, f_{\rm sky} \\
    \times\left[
        \mathrm{Tr}\!\left(
            \hat{\mathbf{C}}_{b}\,
            \mathbf{C}_{b}^{-1}(\theta)
        \right)
        + \ln \lvert \mathbf{C}_{b}(\theta) \rvert
    \right]
    \label{eq:wishart_like}
\end{multline}
where $\hat{\mathbf{C}}_{b}$ and $\mathbf{C}_{b}(\theta)$ denote the measured and model covariance matrices in bin $b$, respectively. We sample this likelihood using an MCMC algorithm and report marginalized constraints on $r$ and $A_{\rm lens}$ from the resulting posterior chains. 

\subsection{Implementation with \texttt{FURAX}} \label{subsec:implementation}

The analysis pipeline is implemented using \texttt{FURAX} 
~\cite{Chanial2026}, an open-source \texttt{Python} framework designed to 
construct and manipulate linear operators for inverse problems in 
astrophysics and cosmology. Built on \texttt{JAX}, \texttt{FURAX} provides a 
modular operator algebra inspired by \texttt{PyOperators}~\cite{Chanial2012} and 
\texttt{lineax}~\cite{rader2023}, enabling an explicit representation of CMB acquisition and 
map-making pipelines. In this framework, the instrumental response is 
expressed as a composition of linear operators describing the optical chain, 
detector response, and scanning strategy, schematically written as
\begin{equation}
    \texttt{P} = \texttt{instrument} \; @ \; \texttt{pointing} \; @ \; \texttt{rotation},
\end{equation}
such that the forward operation $\texttt{d}=\texttt{P(s)}$ corresponds to the data acquisition process, yielding a TOD for a given sky template $\texttt{s}$. The transpose $\texttt{P}^\texttt{T}\texttt{(d)}$ and inverse 
operations are then used in the map-making step, with $\texttt{P}^{-1}$ 
evaluated iteratively using conjugate-gradient solvers. This operator-based formulation allows efficient implementation of  generalized least-squares map-making, while leveraging \texttt{JAX}'s just-in-time compilation, automatic differentiation, and GPU acceleration. 
Compared to existing CMB analysis tools, \texttt{FURAX} enables 
differentiable, end-to-end modeling of instrumental effects and foregrounds 
within a unified framework, making it well suited for systematic studies and 
pipeline prototyping in next-generation CMB experiments.

In the present analysis, the instrumental response is represented, following
Eq.~\ref{eq:Mopticchain},\ref{eq:datamodel1}, by the operator
\begin{equation}
    \texttt{instrument}_{\texttt{X}}
    =
    \texttt{bandpass}
    \; @ \;
    \texttt{polarizer}
    \; @ \;
    \texttt{HWP}_{\texttt{X}},
\end{equation}

\noindent where the index \(\texttt{X}\) labels the assumed HWP model (ideal, effective, stack or non ideal).
\section{Application on a single frequency CMB-only case} \label{sec:appcmb}

In this section we apply the analysis summarized in Fig.~\ref{fig:pipeline} to single frequency CMB-only sky.
As a validation of the pipeline, we verified that in the absence of bandpass 
effects and when the instrumental model used for map-making matches that used to generate the time-ordered data, the input CMB signal is recovered with negligible residuals (see App.~\ref{sec:appendixA}).

We apply the pipeline to a simplified but realistic configuration in order to isolate the impact of bandpass-integrated HWP non-idealities on CMB map reconstruction. The analysis is restricted to a single frequency channel and to simulations containing only a CMB sky signal, with no foreground emission. In all cases, the time-ordered data are generated including bandpass effects and the frequency-dependent non-ideal HWP response, while different HWP models are assumed in the map-making step. As a reminder, the simulated time-ordered data are generated as
\begin{equation}
    \mathbf{d}_{t}(\nu_c)
    =
    \int d\nu\;
    \boldsymbol{\mathcal{B}}(\nu_c, \nu)\,
    \mathbf{P}^{\rm input}_{\rm non id, (t,p)}(\nu)\,
    \mathbf{s}^{\rm CMB}_p(\nu),
    \label{eq:dvalcmb}
\end{equation}

\noindent where $\mathcal{B}(\nu)$ denotes the 
instrumental bandpass and
$\mathbf{P}^{\rm input}_{\rm non id}$ includes the full 
frequency-dependent
instrumental response, modeled using the non-ideal HWP 
described in 
Sec.~\ref{subsec:nonidealhwp}. The map-making step is 
performed employing 
different HWP models of varying complexity 
through the pointing 
matrix $\mathbf{P}^{\rm model}_{\rm X}$, as described in Sec.\ref{subsec:map-making}. Specifically, 
we reconstruct maps 
assuming three different HWP models: the ideal HWP, the effective HWP, and the stacked HWP 
model. The reconstruction using an ideal HWP is taken as 
the baseline configuration, as introduced earlier and corresponds to the uncorrected maps in 
the following. 

\begin{figure}[h]
    \centering
    \includegraphics[width=\columnwidth]  {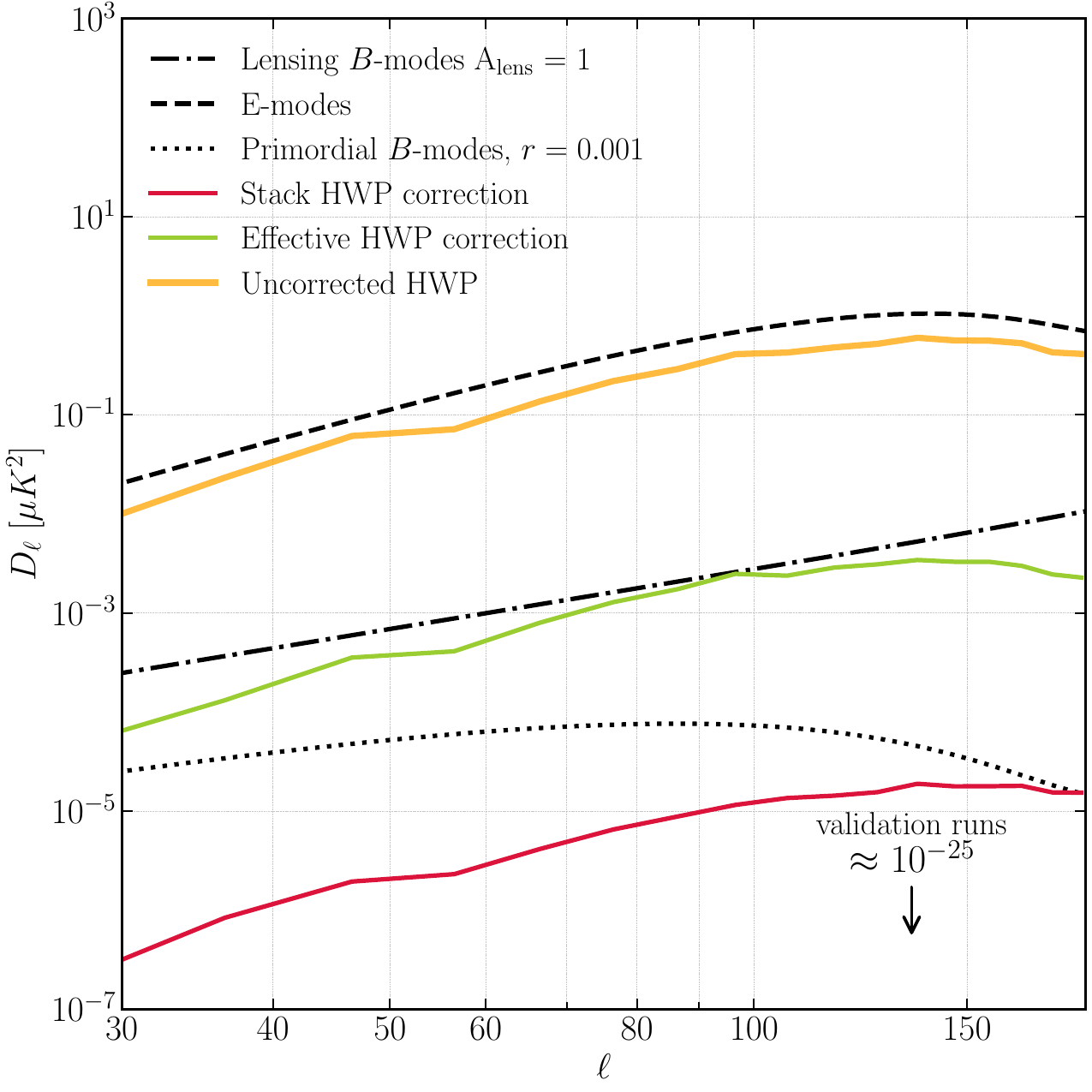}
    \caption{Recovered CMB $B$-modes power spectra 
    residuals for the uncorrected, effective, and stack 
    HWP models for a CMB-only sky. The uncorrected case 
    exhibits significant contamination, whereas both 
    correction schemes substantially reduce the residual 
    bias. The stack correction yields the lowest 
    residual level. These are to be compared with the residual level obtained in the validation runs see Appendix~\ref{sec:appendixA} at $\sim 10^{-25}$ where the input and model 
    HWP descriptions are identical.}
    \label{fig:BBCMB}
\end{figure}

Fig.~\ref{fig:BBCMB} shows the CMB $B$-modes spectrum of the residual map recovered at 150GHz  in \emph{yellow solid line} and compared to the spectra recovered assuming other models, effective and stacked, displayed with \emph{green} and \emph{red lines} respectively. In the absence of HWP correction, significant residual contamination remains across the 
multipole range considered. Implementing the effective 
correction reduces the residual power substantially, 
while the stack correction yields the lowest residual 
spectrum. Quantitatively, the stack model suppresses the residuals by nearly two orders of magnitude 
relative to the effective correction and 
remain well below the lensing $B$-modes signal over the 
full multipole range shown.  This should be compared to the residual level obtained in the validation runs in App.~\ref{sec:appendixA}, which reached down values as low as 10$^{-25}$, and are thus indicative of the numerical accuracy of the method. These results
demonstrate that, even for a frequency-independent CMB
signal, bandpass-integrated HWP non-idealities can
induce significant biases in the recovered $B$-modes
power spectrum if not properly accounted for at the map
making stage.

\section{Application on a multi-frequency CMB+dust case}
\label{sec:appdust}

In this section, we apply the analysis pipeline 
summarized in Fig.~\ref{fig:pipeline} to a multi-frequency sky 
model including both CMB and polarized dust emission. 
This configuration allows us to assess the combined 
impact of bandpass-integrated HWP non-idealities and 
foreground component separation on the recovery of the 
CMB $B$-modes signal.

\subsection{Standard approaches}
\label{subsec:RCC}
We first adopt the standard analysis pipeline detailed in Fig.~\ref{fig:pipeline}. In this approach, instrumental effects are assumed to be fully corrected at the map-making stage (see  Secs.~\ref{subsec:nonidealhwp}-~\ref{subsec:map-making}), and the subsequent component separation is performed using the conventional data model of Eq.~\ref{eq:dAs}. 

\begin{figure*}[ht!]
    \centering
    \includegraphics[width=\textwidth]{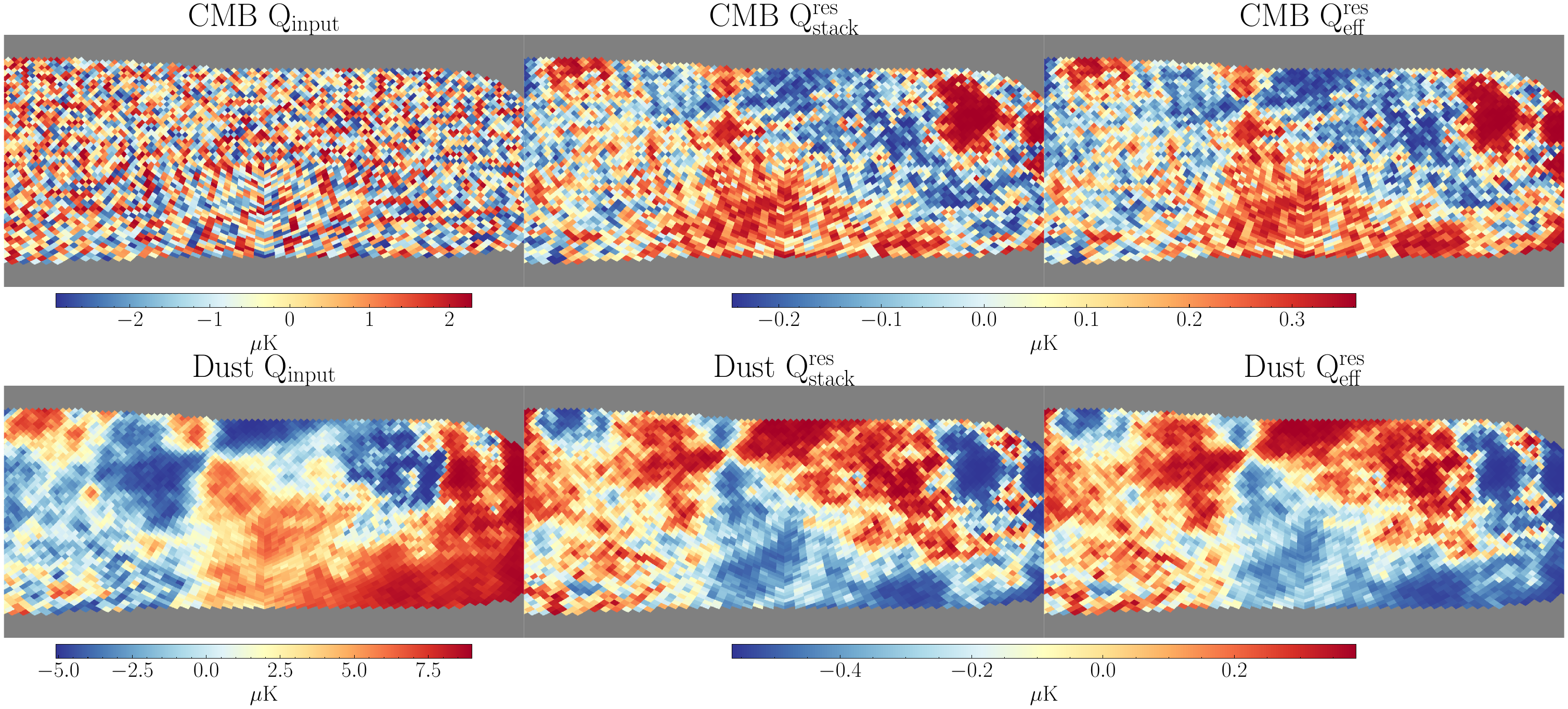}
    \caption{Component separated CMB (\textit{top row}) and dust (\textit{bottom row}) residual maps at $\nu_c= 150$ GHz for each HWP model: effective, and color-corrected stack (from left to right) and the input maps.}
    \label{fig:compsepRCC}
\end{figure*}

We implement a color-corrected stacked method of the map-making model in which the bandpass integration of the instrumental response is performed jointly with the spectral scaling of the foreground emission. Rather than averaging the HWP Mueller matrix elements alone, we use the bandpass-averaged response defined in Eq.~\eqref{eq:stackbpav}. In this expression, the factor $\mathcal{F}(\nu,\nu_c)$ represents the component mixing matrix describing the frequency scaling of the astrophysical emission across the band relative to the central frequency $\nu_c$. We recall that for Galactic dust, we assume a modified blackbody spectrum with spectral index $\beta_d = 1.54$ and temperature $T_d = 20\,\mathrm{K}$, such that $\mathcal{F}(\nu,\nu_c)$ corresponds to the associated dust spectral scaling. This effectively weights the bandpass integration by the dust spectral energy distribution --- in practice, this would be an approximation since the dust SED is expected to be e.g. spatially varying across the sky and along the line-of-sight ---  when constructing the pointing model used in the map-making step.
Fig.~\ref{fig:compsepRCC} shows the component-separated residual maps for the CMB (top row) and 
dust (bottom row) at 150\,GHz for the two HWP 
treatments, color-corrected stacked and effective HWP. Both corrections lead to noticeable large-scale structures in the CMB residuals, but the stack model yields the smallest residuals amplitude overall, substantially suppressing leakages in both CMB and dust maps.

\noindent The results for the recovery of the dust spectral index \(\beta_d\) are summarized in Table~\ref{tab:compsepall}. The quoted uncertainties are obtained from the curvature of the spectral likelihood defined in Eq.~\ref{eq:lspec}. They are purely statistical, as no additional systematic or foreground-modelling uncertainties are included in the likelihood.
For both HWP treatments, the inferred values remain biased with respect to the input, with relative biases of approximately \(5.8\%\) for the effective correction and \(3.9\%\) for the stack correction. These discrepancies are still larger than the \(\sim 1\%\) statistical accuracy typically achieved in current SO component separation pipelines~\cite{Wolz_2024}.

\begin{figure}[ht!]
    \centering
    \includegraphics[width=\columnwidth]{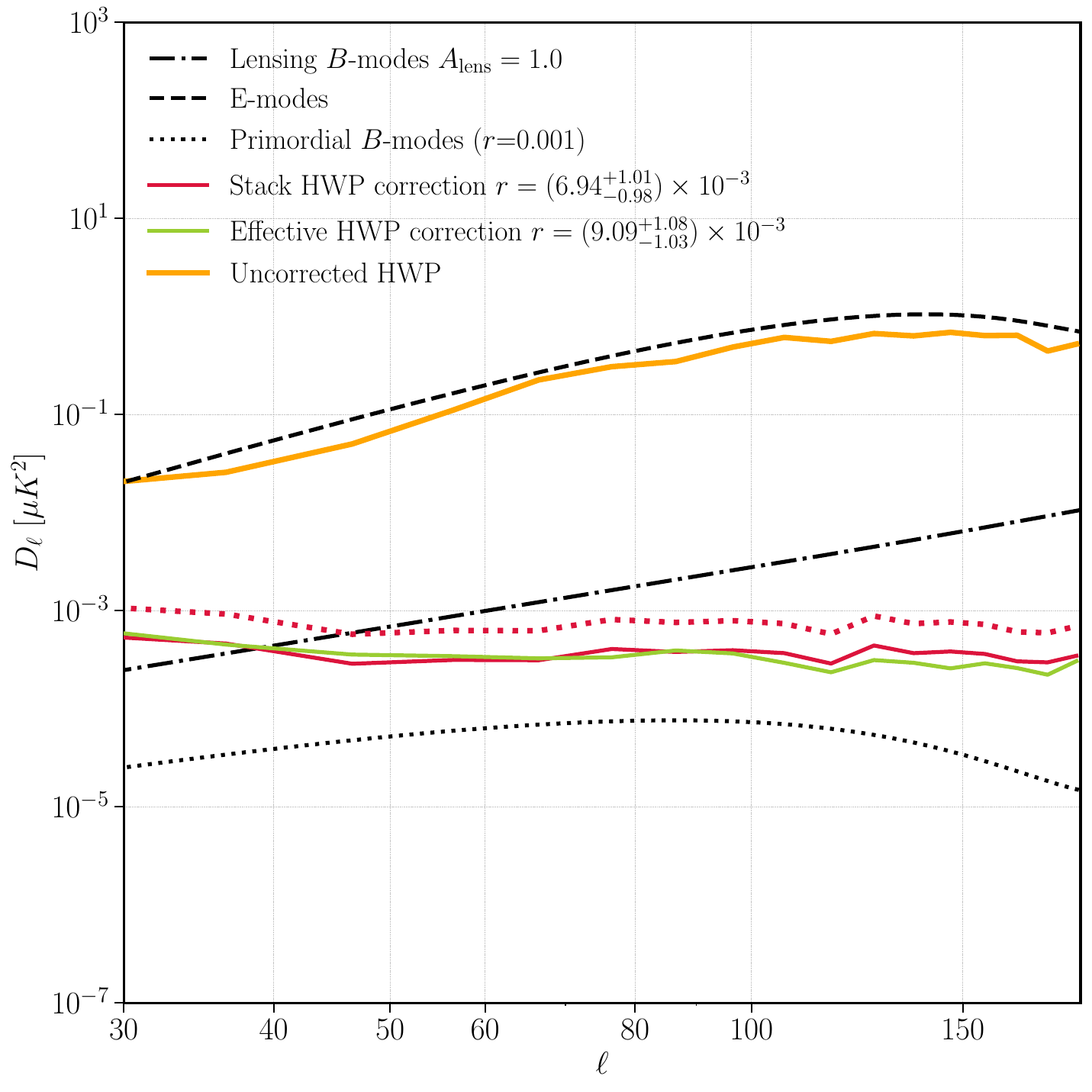}
    \caption{Recovered CMB $B$-modes power spectra after component separation with the stack HWP correction (\textit{red}), the effective HWP correction (\textit{green}), and the uncorrected HWP case (\textit{orange}). The \textit{red dashed} curve represents the residual power spectra when no color correction is applied during map-making for the color-corrected stack model. The stack correction yields the lowest residual contamination, while the uncorrected case exhibits the highest level of residual power across the multipole range.}
    \label{fig:BBRCC}
\end{figure}

In Fig.~\ref{fig:BBRCC}, we present the residual power spectra of the recovered CMB $B$-modes, defined according to Eq.~\ref{eq:rescl}, following component separation. The spectra are shown for three HWP treatments: the stacked HWP correction (with and without color correction), the effective HWP correction, and the uncorrected HWP case. In the following, we consider the color-corrected implementation as the baseline for the "stack correction", as it results in lower residuals in the recovered spectra. The stack HWP correction yields a residual tensor-to-scalar ratio of $r = (9.09^{+1.08}_{-1.03}) \times 10^{-3}$, while the effective HWP correction results in $r = (6.85^{+0.97}_{-0.94}) \times 10^{-3}$ residual. The uncorrected HWP case, however, exhibits the highest level of contamination across the entire multipole range. These findings underscore the importance of precise HWP modeling in mitigating systematic biases, though residual contamination persists at large angular scales due to instrumental mismodeling and limitations in foreground subtraction.

\subsection{Generalised component separation}
\label{subsec:appgencs}

Sec.~\ref{sec:appcmb} and~\cite{verges2021framework} demonstrate that no pointing matrix formulation allows for an exact reconstruction of bandpass-integrated $\mathrm{Q}$ and $\mathrm{U}$ CMB maps, inevitably producing non-negligible residuals in $B$-modes estimation through a standard parametric component separation. These results emphasize that instrumental effects such as HWP frequency dependence cannot be fully  addressed at the map-making level alone. Instead, \emph{these effects must be jointly considered with foreground treatment}. This motivates the  extension of the parameterized component separation method to account for both instrumental systematics and astrophysical emissions, Fig.~\ref{fig:pipeline}.

The mixed-Stokes bandpass averaged maps \(\boldsymbol{\bar{\mathcal{C}}}_4\) and \(\boldsymbol{\bar{\mathcal{S}}}_4\) defined in Eqs.~(\ref{eq:CsampleDef}, \ref{eq:SsampleDef}) are now the input of the component separation. They are recovered at the map-making stage by projecting the data onto the pure \(4f_{\rm HWP}\) harmonics using an ideal-HWP pointing matrix,
\begin{align}
    \begin{pmatrix}
        \boldsymbol{\bar{\mathcal{C}}}_{4p}(\gamma, \nu_c) \\
        \boldsymbol{\bar{\mathcal{S}}}_{4p}(\gamma, \nu_c)
    \end{pmatrix}
    & \equiv
    \mathbf{\hat{s}} \nonumber \\
    & =
    \left(
        \mathbf{P}_{\rm ideal}^T
        \mathbf{P}_{\rm ideal}
    \right)^{-1}
    \mathbf{P}_{\rm ideal}^T \, \mathbf{d}^{\mathrm{non\,id}}_t(\nu_c).
    \label{eq:CS4_application}
\end{align}

As a validation of the pipeline, we verified that when the instrumental model used for map-making matches that used to generate the time ordered data, the input CMB signal is recovered with negligible residuals (see Fig.\ref{fig:BBcompsepval} in App.~\ref{sec:gencsval}).

The component separation stage assumes the simplified instrumental description based on the stack HWP model. We estimate the dust spectral index by maximising the spectral likelihood associated with the generalised component-separation model, obtaining
\(\beta_{d}=1.52 \pm 0.009\), Tab.\ref{tab:compsepall}, where the quoted uncertainties are obtained from the curvature of the corresponding spectral likelihood. This corresponds to a residual bias of 1.3\%\ with respect to the input value, and represents a significant reduction compared to the standard parametric component-separation results reported in Sec.\ref{subsec:RCC}.
 
\begin{table}[ht!]
    \centering
    \renewcommand{\arraystretch}{1.2}
    \begin{tabular}{lc}
        \hline\hline
        \textbf{Foreground parameter} & \(\beta_d\) \\
        \hline
        Input \texttt{d0} & 1.54 \\
        \hline
        \textbf{Standard CS} \\
        \hspace{3mm} Effective HWP & \(1.63 \pm 0.008\) \\
        \hspace{3mm} Stack HWP & \(1.60 \pm 0.009\) \\
        \hline
        \textbf{Generalised CS} \\
        \hspace{3mm} Stack HWP  & \(1.52 \pm 0.009\)    \\
        \hspace{3mm} Validation & \(1.539 \pm 0.009\)    \\
        \hline\hline
    \end{tabular}
    \caption{Estimated values of the dust spectral index \(\beta_d\) for different map-based HWP correction schemes, using simulated CMB + dust sky templates modulated with a non-ideal HWP for noiseless simulated maps.}
    \label{tab:compsepall}
\end{table}

\begin{figure}[ht!]
    \centering
    \includegraphics[width=\columnwidth]{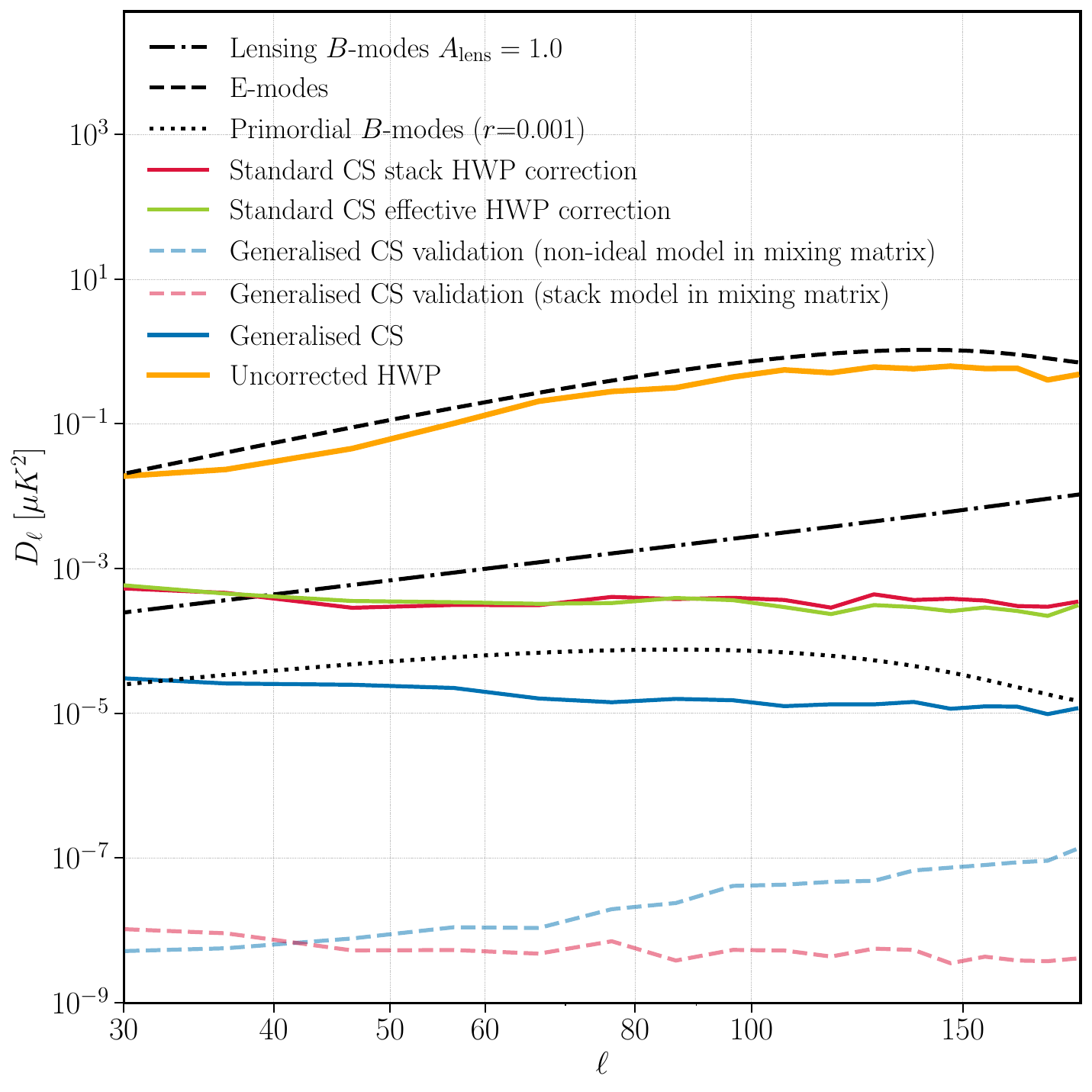}
    \caption{Recovered CMB $B$-modes power spectra for the different correction strategies considered. Solid lines indicate the corresponding residuals with respect to the input CMB signal. Results obtained with standard component separation are shown in \textit{green} for the effective correction and in \textit{red} for the stack correction at the map-making stage, and are compared to the generalised component separation shown in \textit{blue}.}
    \label{fig:BBall}
\end{figure}

Fig.~\ref{fig:BBall} summarizes the performance of all the correction strategies investigated in this work, comparing the residuals at the CMB $B$-modes power spectrum level, shown as colored solid lines, obtained with standard and generalised component separation approaches. When HWP non-idealities are ignored, the reconstructed spectrum exhibits a significant excess of $B$-modes power, with correspondingly large residuals. Successive corrections based on effective and stack HWP models substantially reduce the residuals, but significant large-scale residuals remain when instrumental effects are treated only at the map-making level. In contrast, the generalised component separation, shown as the \emph{blue solid curve}, which explicitly incorporates the instrumental mixing, yields the lowest residual levels, bringing the recovered spectrum close to the input CMB signal across the multipole range considered. We note, however, that the lowest residual levels are obtained in the validation configurations (\emph{dashed colored lines}), where the instrumental response assumed in the recovery matches the HWP model used to generate the simulations perfectly. This illustrates that the performance of the generalised component separation is ultimately limited by the fidelity of the instrumental model entering the mixing matrix. In practice, reaching this level of mitigation therefore would require accurate pre-deployment characterisation and in-flight/in-the-field calibration of the HWP frequency-dependent response and associated leakage terms.

\begin{figure}[ht!]
    \centering
    \includegraphics[width=\columnwidth]{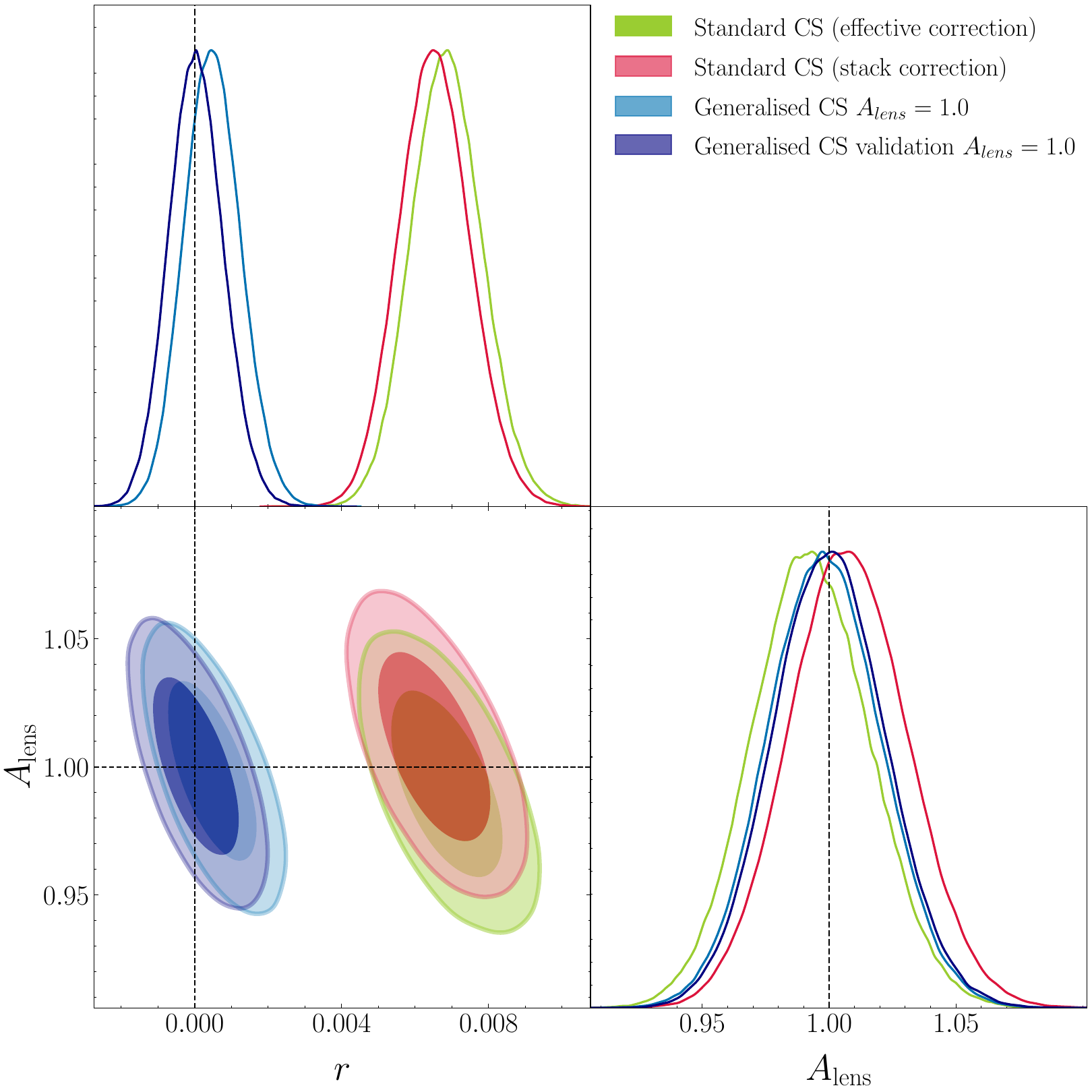}
    \caption{Marginalized and joint posteriors for $(r, A_{\rm lens})$ obtained for the different correction strategies: standard component separation with effective (\textit{green}) and stack (\textit{red}) HWP corrections, and generalised component separation including the stack model in the mixing matrix (\textit{blue}). Dashed lines indicate the input values $r=0$ and $A_{\rm lens}=1$.}

    \label{fig:rAlens}
\end{figure}

Fig.~\ref{fig:rAlens} shows the resulting marginalized posteriors in the \((r, A_{\rm lens})\) space for the different correction treatments. Tab.~\ref{tab:r_Alens_constraints} summarizes the estimates of \(r\) and \(A_{\mathrm{lens}}\) for each treatment. We recall that the data used in this analysis are noiseless, so the reported error bars should be interpreted as lower bounds. The standard component separation results obtained with the effective and stack corrections both yield posteriors for \(r\) that are significantly shifted away from the input value, indicating a clear bias induced by residual systematics. In contrast, the generalised component separation substantially reduces this bias, with the recovered \(r\) consistent with the input value within uncertainties.

\begin{table}[ht!]
\centering

\renewcommand{\arraystretch}{1.5}
\begin{tabular}{lcc}
\hline\hline
& \shortstack{$r$}
& \shortstack{$A_{\rm lens}$ } \\
\hline
\textbf{Input} & $0.0$ & $1.0$ \\
\hline
\textbf{Standard CS} & & \\
\hline
\hspace{3mm} Effective HWP 
& $(9.08^{+1.08}_{-1.03}) \times 10^{-3}$
& $1.02 \pm 0.02$ \\

\hspace{3mm} Stack HWP
& $(6.94^{+1.01}_{-0.98}) \times 10^{-3}$
& $1.02 \pm 0.02$ \\
\hline
\textbf{Generalised CS}
& $(4.80^{+7.70}_{-7.50}) \times 10^{-4}$
& $0.99 \pm 0.02$ \\
\hspace{3mm} Validation
& $(0.02^{+7.6}_{-7.3}) \times 10^{-4}$
& $1.00 \pm 0.02$ \\
\hline\hline
\end{tabular}
\caption{Summary of marginalized noiseless constraints on the tensor-to-scalar ratio $r$ and the lensing amplitude $A_{\rm lens}$ for the different correction strategies. Quoted uncertainties correspond to the posterior median with $68\%$ confidence intervals.}
\label{tab:r_Alens_constraints}
\end{table}

\noindent We further note that the smallest residual bias is obtained in the validation case, highlighting that the accuracy of the recovered constraints ultimately depends on the fidelity of the HWP characterisation entering the generalised mixing model. 

We next extend the analysis by considering an artificially delensed configuration in which the lensing amplitude is fixed to $A_{\rm lens}=0.3$ \cite{hertig2024}. As expected, reducing the lensing contribution narrows the posterior on $r$, reflecting the smaller lensing-induced variance in the $B$-modes spectrum. However, the inferred value of $r$ remains biased with respect to the input value, the reduction of $\sigma_r$ makes the residual bias comparatively more significant, increasing the bias-to-$\sigma_r$ ratio. As shown by the validation case, mitigating this bias requires either an instrumental model that matches the true HWP response to high accuracy, or alternative analysis strategies that incorporate the instrumental response more directly, for instance by performing component separation at the time-domain level.

\begin{figure}
    \centering
    \includegraphics[width=\columnwidth]{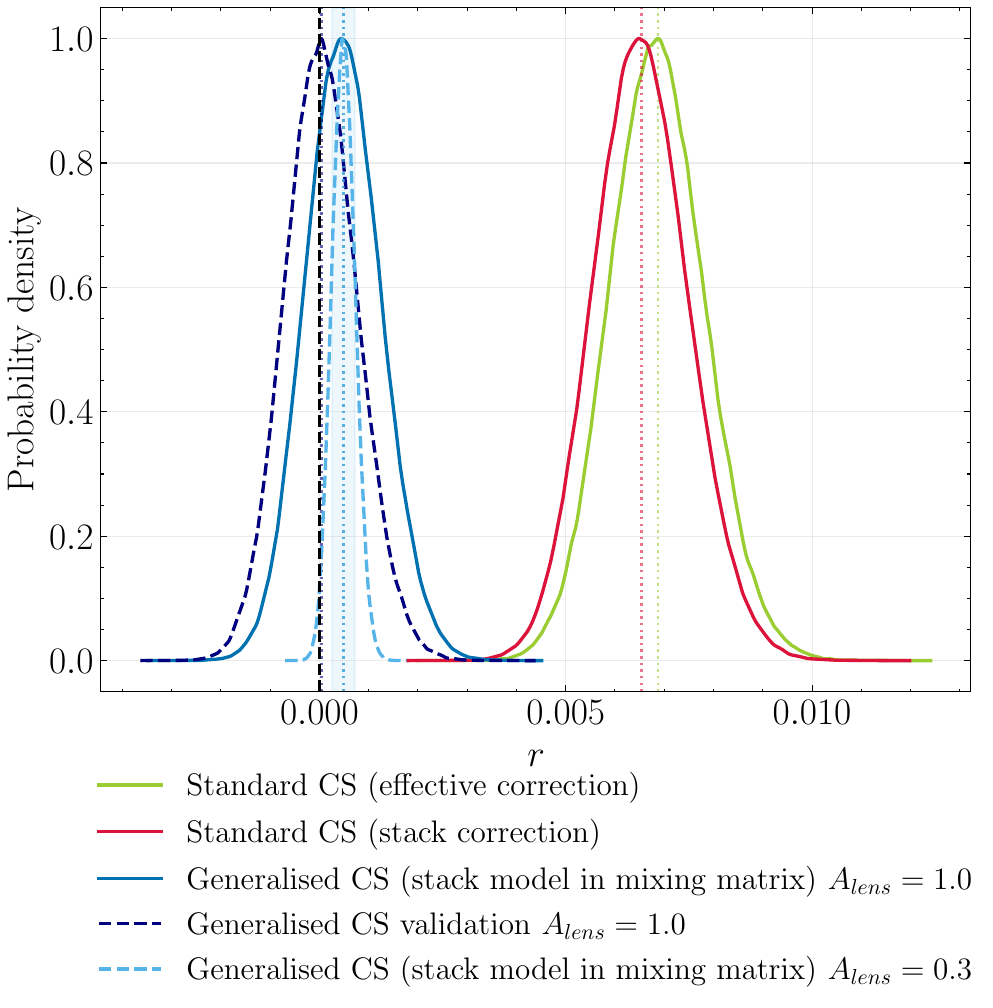}
    \caption{Marginalized posterior distributions for the tensor-to-scalar ratio $r$ for the different correction strategies. The solid curves correspond to the standard component separation with effective (\textit{green}) and stack (\textit{red}) HWP corrections, and to the generalised component separation including the stack model in the mixing matrix (\textit{blue}) assuming $A_{\rm lens}=1.0$. The dashed blue curve shows the corresponding generalised component separation result when fixing $A_{\rm lens}=0.3$, illustrating the reduction in $\sigma_r$ when the lensing contribution is suppressed.}
    \label{fig:restimation03}
\end{figure}

\subsection{Exploration of time-domain component separation}
\label{subsec:tdcompsep}

The generalized component-separation framework developed above operates at the map level and captures the leading impact of frequency-dependent HWP-induced Stokes mixing. Nevertheless, a fully consistent treatment of instrumental effects would ideally be performed directly in the time domain. Working with time-ordered data enables the incorporation of more realistic noise models, including correlated and non-stationary contributions, as well as accounting for detector-dependent characteristics. In ground-based observations, atmospheric emission introduces strong low-frequency noise and correlated structures across detectors, which are typically mitigated through filtering and scan-synchronous subtraction. These operations can couple non-trivially to HWP harmonics (including additional terms arising from non-idealities) and to the scanning strategy, potentially inducing residual leakage that is challenging to capture in a purely map-based model. This is particularly relevant when multiple effects combine, for instance frequency-dependent phase shifts and intensity-to-polarization leakage interacting with time-domain filtering and atmospheric fluctuations. We defer a full time-domain component separation treatment, including a realistic atmosphere model and other systematics, to a future work (\cite{Beringue2025, Errard2026}).

In this section, we retain the same simulated data configuration as in the previous sections and use identical HWP models in both the forward simulation and the separation stage. The purpose is therefore not to provide a complete time-domain treatment, but to demonstrate the feasibility of extending the component-separation framework to operate directly on time-ordered data within a controlled setting.

The data model follows Eq.~\ref{eq:datamodel1} and includes the color correction of the dust emission,
\begin{equation}
\mathbf{d}_t(\nu_c) =
\underbrace{\int d\nu\;
\boldsymbol{\mathcal{B}}(\nu_c,\nu)\,
\mathbf{P}^{\rm input}_{\rm non\,id}(\nu)\,
\mathbf{A}(\nu,\nu_0)}_{\equiv\,\boldsymbol{\mathcal{P}}^{\rm non\,id}(\nu,\gamma)}
\, \mathbf{s}(\nu_0).
\label{eq:data_model_TOD}
\end{equation}

\noindent Here $s(\nu_0)$ denotes the sky templates evaluated at the reference frequency $\nu_0$, while $\mathbf{P}^{\rm input}_{\rm non\,id}$ is the pointing matrix including the non-ideal HWP parameters $\gamma$, which encodes the scanning strategy and detector orientation. The resulting TODs are used as input to the component-separation analysis. For noiseless data, the spectral likelihood can then be written as
\begin{equation}
-2 \log \mathcal{L}_{\rm spec}(\beta) =
\mathbf{\hat{d}}_t^{T}\,\boldsymbol{\mathcal{P}}\,
(\boldsymbol{\mathcal{P}}^{T}\boldsymbol{\mathcal{P}})^{-1}
\boldsymbol{\mathcal{P}}^{T}\mathbf{\hat{d}}_t^{T} ,
\label{eq:lspectcs}
\end{equation}
where, for clarity of notation, the explicit frequency dependence has been omitted.
We estimate the dust spectral index by maximizing the spectral likelihood, obtaining
 \(\beta_{d}=1.54 \pm 0.005\), where the uncertainty is derived from the curvature of the spectral likelihood at its maximum. The likelihood in Eq.~\ref{eq:lspectcs} is evaluated using $10\,\mathrm{s}$ of simulated TOD, corresponding to $N_{\mathrm{sampling}}=10^{4}$ samples per detector, for (20 $\times$ 20) grid detectors. For the scanning strategy considered here, this results in an average number of observations per pixel of $\langle N_{\mathrm{hits}} \rangle \simeq 80$, ensuring a well-sampled sky coverage. Using this best-fit value, we compute TOD residuals for each recovered component, defined as the difference between the sampled input maps and the reconstructed CMB and dust signals. 

\begin{figure}[ht!]
    \centering
    \includegraphics[width=\columnwidth]{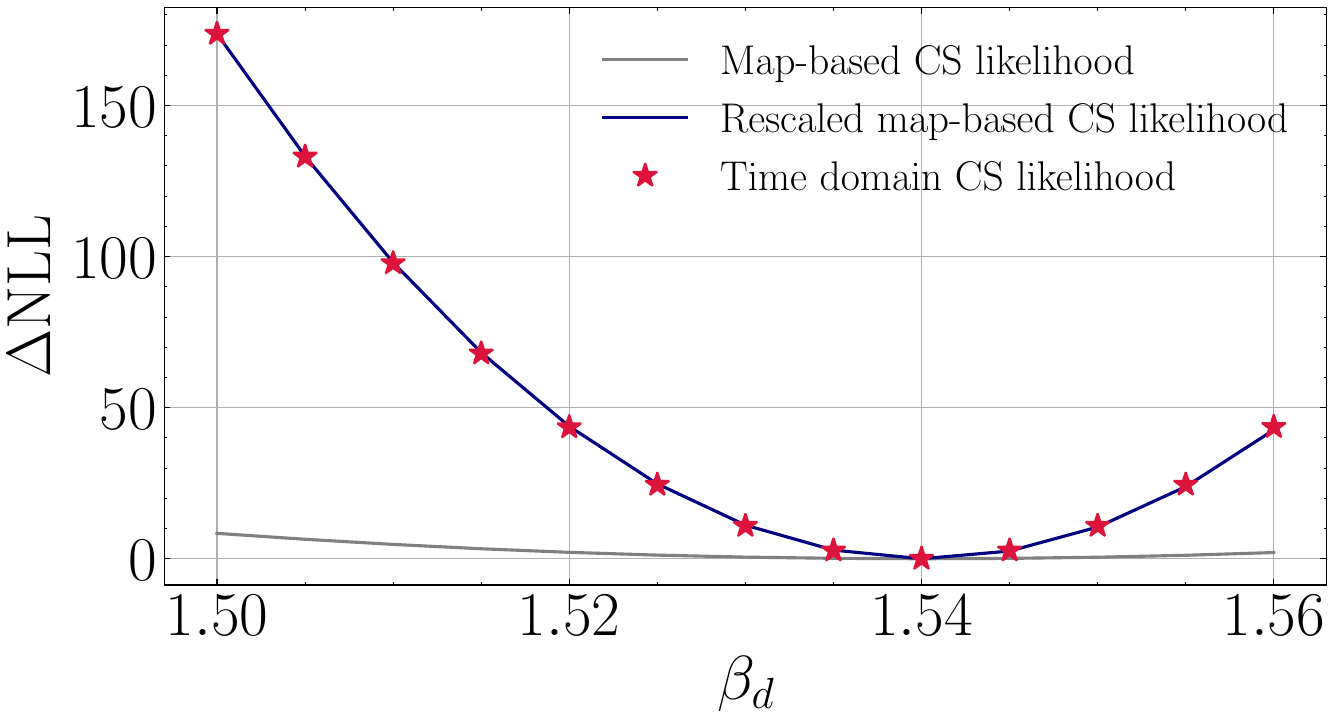}
\caption{
Spectral likelihood as a function of the dust spectral index $\beta_d$. The \emph{gray} curve shows the likelihood computed from pixelized maps assuming uniform noise per pixel. The \emph{blue} curve shows the map-domain likelihood after rescaling the noise covariance by the number of observations per pixel, $N_{\mathrm{hits}}$. The \emph{red markers} show the likelihood computed directly from the time-ordered data (TOD). After the $N_{\mathrm{hits}}$ rescaling, the map-domain and TOD likelihoods become consistent.
}
\label{fig:likelihoodcomp}
\end{figure}

To assess the consistency between the map-domain and time-domain implementations, we compute the spectral likelihood using both pixelized maps and TOD, adopting the same non-ideal HWP model in both the forward simulation and the component-separation step, as in a validation configuration. In the map-domain case, the likelihood computed assumes uniform noise per pixel and therefore does not account for the different number of time samples contributing to each pixel. To ensure a consistent comparison with the TOD analysis, we rescale the map-domain noise covariance by the number of observations per pixel, $N_{\mathrm{hits}}(p)$, which accounts for the averaging of time samples in the map-making process. This rescaling restores the correct statistical weighting of each pixel. As shown in Fig.~\ref{fig:likelihoodcomp}, the rescaled map-domain likelihood becomes statistically equivalent to the likelihood computed directly from the TOD, yielding consistent constraints on $\beta_d$.

\begin{figure}[ht!]
    \centering
    \includegraphics[width=\columnwidth]{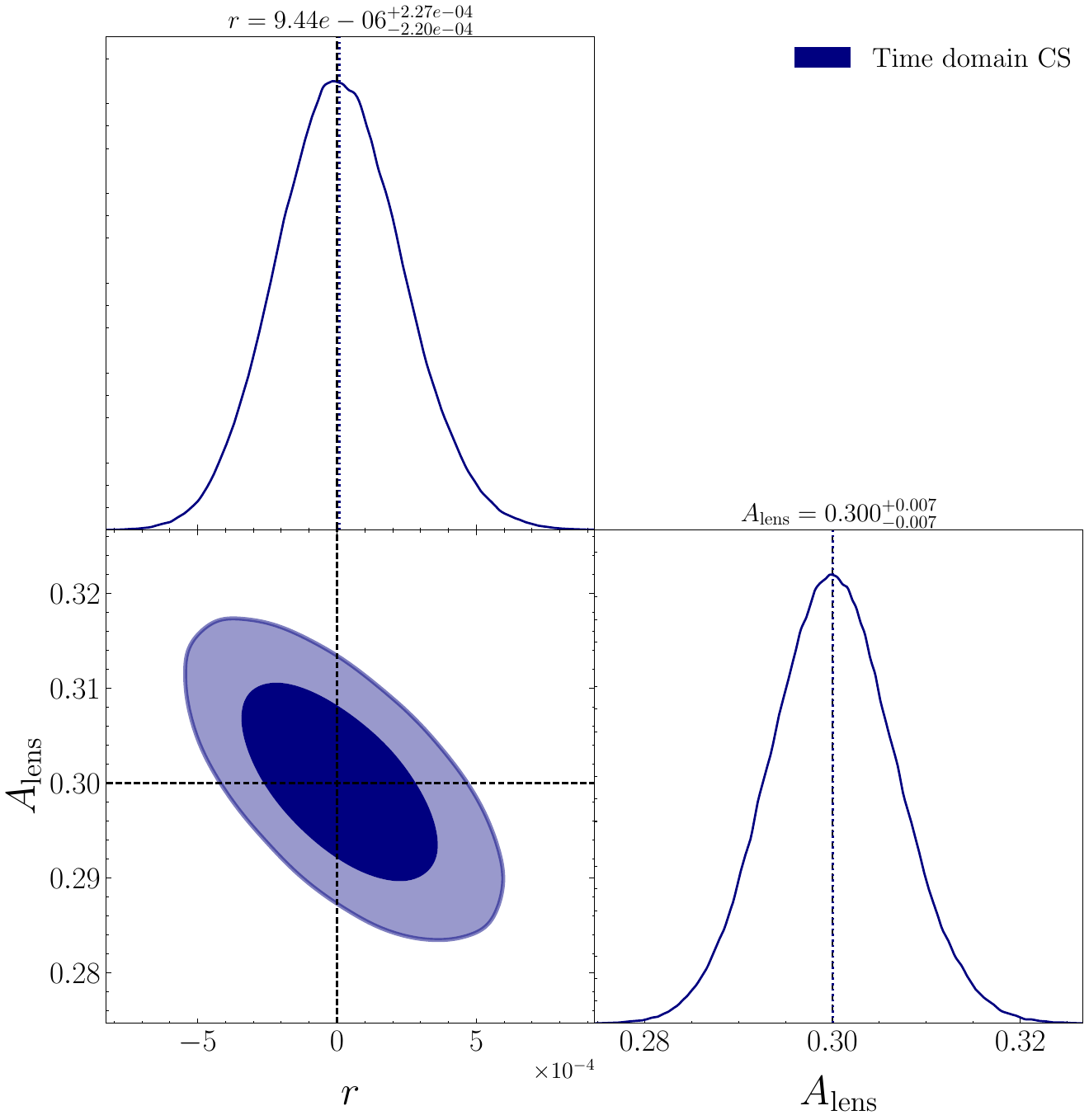}
    \caption{
Marginalized and joint posterior distributions for $(r, A_{\rm lens})$ obtained from the time-domain component-separation pipeline. The recovered parameters are consistent with the input values, yielding $r=(0.09^{+2.27}_{-2.20})\times10^{-4}$ and $A_{\rm lens}=0.300^{+0.007}_{-0.007}$. The constraint on $r$ is consistent with zero well within $1\sigma$.
}
    \label{fig:restimationtcs}
\end{figure}

We consider a delensed configuration with input value $A_{\rm lens}=0.3$. Fig.~\ref{fig:restimationtcs} shows the marginalized posterior distribution in the $(r, A_{\rm lens})$ plane. The recovered parameters are consistent with the input values. In particular, we find $r = (0.09^{+2.27}_{-2.20}) \times 10^{-4}$, corresponding to a value consistent with $r=0$ within $1\sigma$.

Overall, these results demonstrate that the component-
separation framework can be consistently extended to 
operate directly on time-ordered data. In the simplified 
noiseless configuration considered here, the recovered 
spectral parameters and cosmological constraints are 
consistent with the input values, and the likelihood 
computed from TOD agrees with the properly rescaled map-domain likelihood. This validates the map-based 
formalism developed in the previous sections while 
confirming that the underlying methodology naturally 
generalizes to a full time-domain treatment. 

We also note that the computational cost of this approach remains tractable for simplified instrumental models: for the simplest HWP configuration (ideal or stack HWP) considered here, the full analysis can be performed on timescales of order minutes. For the more complex HWP description, the runtime increases to $\mathcal{O}(10)$ minutes, with the dominant cost arising from evaluations of the HWP model rather than the minimization procedure itself. The computational cost will increase for more realistic instrumental models and larger data volumes. However, the problem is well suited to parallelization and hardware acceleration and promising relevant numerical algorithms have already been proposed and studied~\cite{Papez2020}.  
\section{Conclusions} \label{sec:Conclusion}

In this work, we investigated the impact of frequency-dependent HWP non-idealities on CMB polarization measurements, in the presence of galactic foregrounds and lensing. Starting from a   a Mueller matrix description of the full optical chain, we showed that realistic multi-layer HWPs introduce a frequency-dependent mixing of the Stokes parameters, breaking the commonly assumed simple $4f_{\rm HWP}$ modulation of $\mathrm{Q}$ and $\mathrm{U}$ input signals.

Using end-to-end simulations including bandpass integration, we quantified how these effects propagate through map-making and standard parametric, map-based component separation. We find that the resulting mismodelling produces non-negligible residuals in the recovered CMB maps and in the $B$-mode power spectrum, particularly on large angular scales.

We proposed and tested map-level mitigation strategies based on effective and stacked HWP corrections. While these approaches substantially reduce the contamination, we find they do not suppress it to the level required to meet the science goals of current-generation CMB polarization experiments, i.e. $r\lesssim \mathcal{O}(0.001)$.

Motivated by these limitations, we developed a generalized component-separation framework that explicitly incorporates instrumental parameters into the mixing matrix. This formulation enables a joint treatment of instrumental response and astrophysical emission. In controlled validation tests, the method yields unbiased recovery of both the CMB $B$-mode signal and dust spectral parameters. In more realistic scenarios, where mismatches between simulated and modelled data are present, it provides a significant additional reduction of residual contamination compared to standard component-separation approaches. The recovered values of $r$ are consistent with the input within uncertainties. We also considered an artificially delensed configuration with reduced lensing amplitude. Although delensing decreases the statistical uncertainty on $r$, residual biases arising from foreground and instrumental mismodelling remain and become comparatively more important as $\sigma_r$ decreases.

These results suggest extending component separation to more accurate and flexible descriptions of the HWP response. In particular, a consistent treatment would involve jointly fitting instrumental parameters together with foreground spectral parameters. Ultimately, this points toward performing component separation directly in the time domain, where frequency-dependent instrumental effects and time-dependent systematics can be treated within a unified framework. We demonstrated the feasibility and validation of such an approach with a consistent estimation $\beta_d$ with respect to the pixel domain method and of $r$ with respect to the input values. 

Our results demonstrate that frequency-dependent HWP-induced mixing of Q and U represents a fundamental limitation for standard analysis frameworks, as it challenges their core assumptions and requires a revision of the underlying data modeling.  Developments along the lines above  appear to be necessary to achieve robust measurements of primordial $B$-modes at the sensitivity targets of forthcoming CMB experiments and to ensure unbiased recovery of cosmological parameters.
\section*{Acknowledgements}

The authors would like to thank the \textsc{SciPol} team, Clément Leloup, Tomotake Matsumura, Nicòlo Raffuzzi and Clara Vergès for useful discussions and feedback during the project.

This work was carried out within the \textsc{SciPol} project (\href{https://scipol.in2p3.fr}{scipol.in2p3.fr}), supported by the European Research Council (ERC) under the European Union’s Horizon 2020 research and innovation program (Grant Agreement No.~101044073, PI: J. Errard). 

Computations were performed on the \textit{Jean Zay} supercomputer at IDRIS, using HPC resources provided by GENCI under allocations 2024-AD010414161R2 and 2025-A0190416919. 

The authors acknowledge financial support from the Centre national d’études spatiales (CNES), France (ROR: \href{https://ror.org/04h1h0y33}{https://ror.org/04h1h0y33}), within the framework of the LiteBIRD space mission.

This work has received funding by the European Union’s Horizon 2020 research and innovation program under grant agreement no. 101007633 \emph{CMB-Inflate}.

The authors acknowledge support from the France-Berkeley Fund through the project “Instrumental systematics for the next generation of Cosmic Microwave Background experiments” (PIs: C.~Vergès and J.~Errard).
\appendix
\appendix

\section{Amplitude--phase representation of the polarization modulation}
\label{app:amp_phase}

\subsection{Relation to the HWP Mueller matrix}

We detail how the expressions of the $\mathbf{C_{0i;k}}(\nu), \mathbf{S_{0i;k}}(\nu)$ coefficients can be written directly in terms of the elements of the
frequency--dependent HWP Mueller matrix $\mathbf{H}(\nu)$. For the $4f$ harmonic, $k=4$, one obtains
\begin{align}
\mathbf{C}_{01;4}(\nu) &= \frac12\left(\mathbf{H}_{11}(\nu)-\mathbf{H}_{22}(\nu)\right),\\
\mathbf{S}_{01;4}(\nu) &= -\frac12\left(\mathbf{H}_{12}(\nu)+\mathbf{H}_{21}(\nu)\right),\\
\mathbf{C}_{02;4}(\nu) &= -\frac12\left(\mathbf{H}_{12}(\nu)+\mathbf{H}_{21}(\nu)\right),\\
\mathbf{S}_{02;4}(\nu) &= -\frac12\left(\mathbf{H}_{11}(\nu)-\mathbf{H}_{22}(\nu)\right).
\end{align}

\subsection{Amplitude--phase form}

We present the relation between bandpass averaged $\bar{\mathbf{C}}_{0i;k}(\nu), \bar{\mathbf{S}}_{0i;k}(\nu)$ coefficients and the amplitude and phase coefficients appearing in description of the stacked HWP model in Sec.\ref{subsubsec:mapmakingstack}. We note,

\begin{equation}
\theta_t = 4\varphi_t + 2\alpha_t .
\end{equation}
Any linear combination of sine and cosine functions with the same phase can
be written as a single phase--shifted sinusoid,
\begin{equation}
A\cos\theta + B\sin\theta
=
R\cos(\theta+\psi),
\end{equation}
where
\begin{equation}
R = \sqrt{A^2+B^2},
\qquad
\psi = \tan^{-1}\left(\frac{B}{A}\right).
\end{equation}

\noindent Applying this identity to the $\mathrm{Q}_t$ terms in Eq.~(\ref{eq:fitstack}) yields
\begin{multline}
\mathrm{Q}_t\left[
\bar{\mathbf{C}}_{01,k}^{\mathrm{stack}}\cos\theta_t
+
\bar{\mathbf{S}}_{01,k}^{\mathrm{stack}}\sin\theta_t
\right] \\
=
\mathrm{Q}_t\,\eta_Q(\nu_c)\cos(\theta_t+\psi_Q(\nu_c)),
\end{multline}
with
\begin{align}
\eta_Q(\nu_c) &=
\sqrt{
\left(\bar{\mathbf{C}}_{01,k}^{\mathrm{stack}}\right)^2
+
\left(\bar{\mathbf{S}}_{01,k}^{\mathrm{stack}}\right)^2},\\
\psi_Q(\nu_c) &=
\tan^{-1}\left(
\frac{\bar{\mathbf{S}}_{01,k}^{\mathrm{stack}}}
{\bar{\mathbf{C}}_{01,k}^{\mathrm{stack}}}
\right).
\end{align}

\noindent An identical procedure applied to the $\mathbf{U}_t$ terms gives
\begin{multline}
\mathrm{U}_t\left[
\bar{\mathbf{C}}_{02,k}^{\mathrm{stack}}\cos\theta_t
+
\bar{\mathbf{S}}_{02,k}^{\mathrm{stack}}\sin\theta_t
\right] \\
=
\mathrm{U}_t\,\eta_U(\nu_c)\sin(\theta_t+\psi_U(\nu_c)),
\end{multline}
where
\begin{align}
\eta_U(\nu_c) &=
\sqrt{
\left(\bar{\mathbf{C}}_{02,k}^{\mathrm{stack}}\right)^2
+
\left(\bar{\mathbf{S}}_{02,k}^{\mathrm{stack}}\right)^2},\\
\psi_U(\nu_c) &=
\tan^{-1}\left(
\frac{\bar{\mathbf{S}}_{02,k}^{\mathrm{stack}}}
{\bar{\mathbf{C}}_{02,k}^{\mathrm{stack}}}
\right).
\end{align}

\section{Standard component separation validation} \label{sec:appendixA}

To validate our pipeline end-to-end, we consider a single frequency channel 
and neglect bandpass effects. The simulations include only the CMB sky 
signal, with no foreground emission. The time-ordered data are generated by 
modulating the input sky signal 
with a given HWP model at the selected frequency from Eq.~\ref{eq:datamodel1}. 
The same model X is  used to construct the pointing matrix $ \mathbf{P}^{\rm model}_{\rm X}$ in the map-making step 
Eq.~\ref{eq:mapmakingsolbis} ---  i.e. $\mathbf{P}^{\rm model}_{\rm X} = 
\mathbf{P}^{\rm input}_\texttt{X}$ in this paragraph. This 
validation procedure is applied to the two HWP corrections considered in 
this work, corresponding to $\mathrm{X}\,\in\,\{$effective, 
stack$\}$. To quantify the reconstruction accuracy, we compute residual maps 
defined in Eq.~\ref{eq:resmap} for a given HWP model, and compute the 
corresponding residual $B$-modes power spectrum defined in 
Eq.~\ref{eq:rescl}.  For all HWP models, the residual $B$-modes power 
spectra are $<10^{-25}\,\mu{\rm K}^2$, i.e. consistent with zero, 
demonstrating that the map-making procedure accurately reconstructs 
the input sky signal when the instrumental model is correctly specified as 
we expect. 

\begin{figure}[ht!]
    \centering
    \includegraphics[width=\columnwidth]{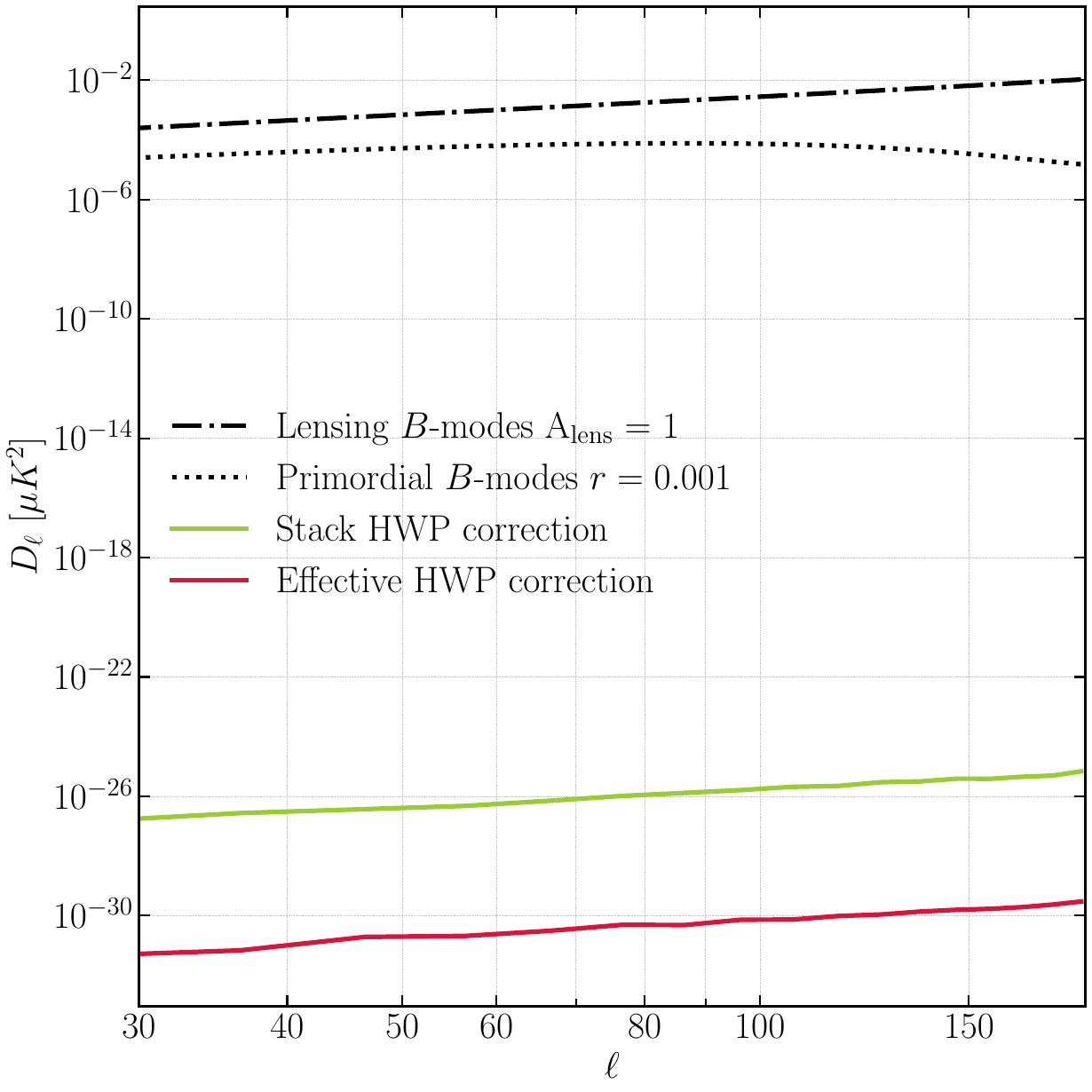}
    \caption{Recovered CMB residual $B$-modes power spectra for each HWP model. Residuals are effectively zero compared to the lensing $B$-modes and primordial $B$-modes, confirming out map-making procedure.}
    \label{fig:BBCMB_validation}
\end{figure}

\section{Generalised component separation validation}
\label{sec:gencsval}

In this section we validate the generalised component separation framework by considering cases in which the data model assumed in the component separation matches the one used to generate the time-domain simulations. Exploiting the time-domain structure of Eq.~\ref{eq:stack_model}, we recover the mixed-Stokes maps by performing map-making with an ideal-HWP in the pointing matrix Eq.~\ref{eq:CS4_application} and thus directly projects the data onto the mixed-Stokes basis. We include the instrumental response into a generalised mixing matrix as described in Sec.~\ref{subsubsection:gencs}. This validation procedure is applied both to the stack HWP model and to the fully non-ideal HWP model used in the simulations. These two cases allow us to assess the mitigation performance of the generalised component separation in a controlled setting: the stack model provides a compact description that captures the dominant frequency dependence of the HWP response, while the fully non-ideal model represents the most complete instrumental response used to generate the data.
We do not consider the effective model here, as it is primarily intended as a simplified map-level correction and does not capture the full frequency-dependent mixing that the generalised component separation is designed to mitigate. The corresponding estimates of the dust spectral index are reported in Table~\ref{tab:gencompsepval}, the quoted uncertainties are obtained from the curvature of the corresponding spectral likelihood. In both validation configurations, the recovered \(\beta_d\) values are consistent with the input, demonstrating unbiased spectral-parameter recovery when the component-separation model matches the instrumental response used in the simulations.

\begin{table}[ht!]
    \centering
    \begin{tabular}{lc} 
        \hline\hline
        \textbf{Foreground parameter} & \(\beta_d\) \\ 
        \hline
        Input \texttt{d0} & 1.54  \\  
        Stack HWP correction & 1.539 $\pm$ 0.0092 \\ 
        Non ideal HWP & 1.539 $\pm$ 0.0097 \\ 
        \hline\hline
    \end{tabular}
    \caption{Estimated values of the dust spectral index \(\beta_d\) for generalised component separation valisation.}
    \label{tab:gencompsepval}
\end{table}

\begin{figure}[ht!]
    \centering
    \includegraphics[width=\columnwidth]{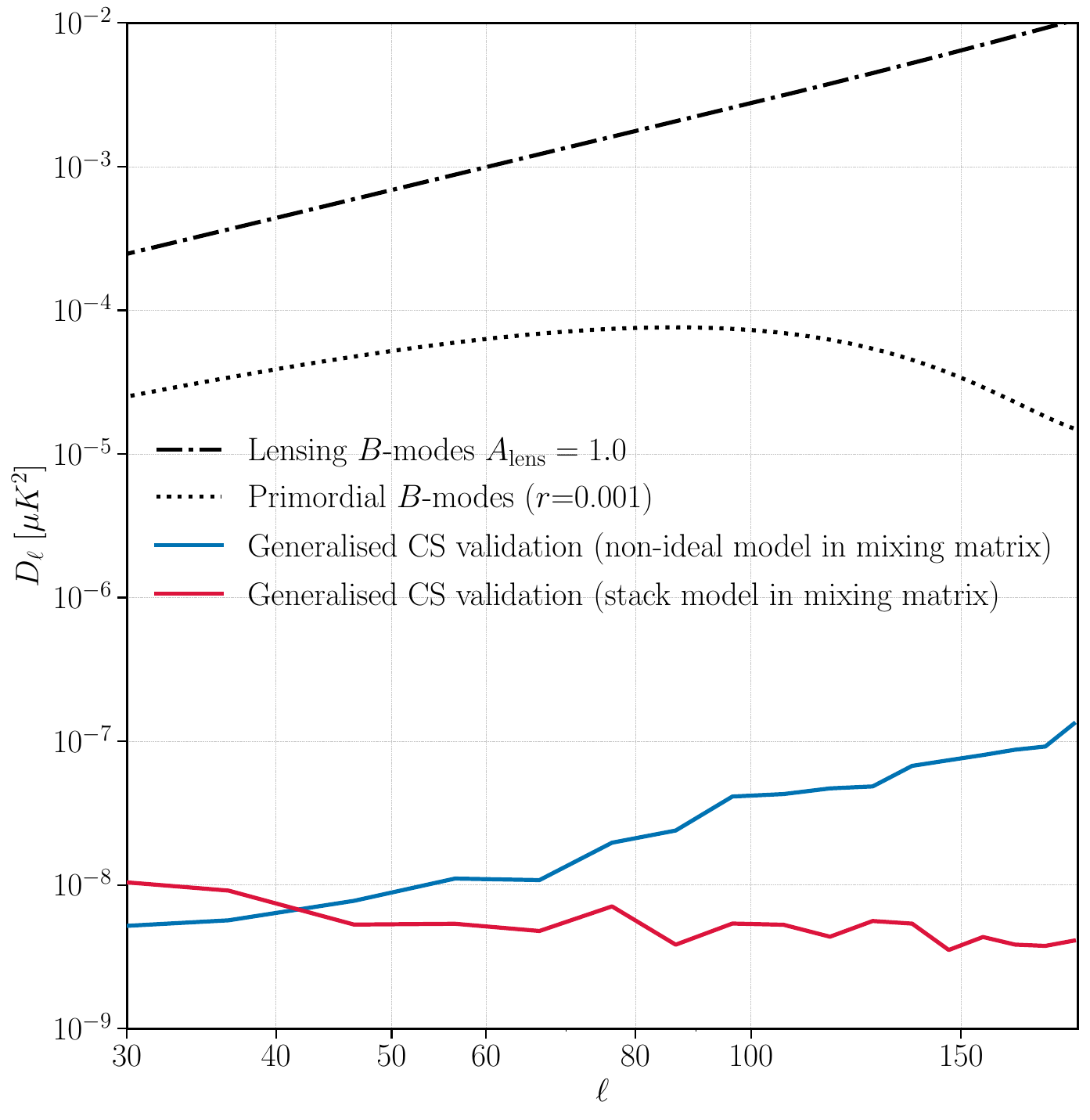}
    \caption{Residual CMB $B$-modes power spectra obtained in the validation configuration of the generalised component separation framework. The curves show the residual spectra, both the fully non-ideal HWP model (\emph{blue}) and the stack HWP model (\emph{red}) yield residuals well below the target primordial signal level, demonstrating unbiased CMB recovery in the absence of modelling mismatch.
    }
    \label{fig:BBcompsepval}
\end{figure}

Fig.~\ref{fig:BBcompsepval} shows the residual CMB $B$-modes power spectra obtained after applying the generalised component separation in the validation configuration. The solid curves correspond to the residual spectra, \(C_{\ell,\mathrm{res}}^{BB}\), defined following Eq. \ref{eq:rescl}. The residual amplitude remains well below both the lensing signal and a primordial contribution with \(r=10^{-3}\) across the full multipole range considered, $r \approx 2 \times 10^{-6}$ for both models. It demonstrates that the generalised component separation accurately accounts for the frequency-dependent instrumental mixing when the data model is consistent. This validation serves as a benchmark for the analysis scenarios considered, in which the assumed instrumental model differs from the true response present in the data.

\nocite{*}
\bibliography{hwpbib}% Produces the bibliography via BibTeX.

@article{Giardiello_2023,
   title={Detailed study of HWP non-idealities and their impact on future measurements of CMB polarization anisotropies from space (Corrigendum)},
   volume={671},
   ISSN={1432-0746},
   url={http://dx.doi.org/10.1051/0004-6361/202141619e},
   DOI={10.1051/0004-6361/202141619e},
   journal={Astronomy \&; Astrophysics},
   publisher={EDP Sciences},
   author={Giardiello, S. and Gerbino, M. and Pagano, L. and Errard, J. and Gruppuso, A. and Ishino, H. and Lattanzi, M. and Natoli, P. and Patanchon, G. and Piacentini, F. and Pisano, G.},
   year={2023},
   month=mar, pages={C1} }

@misc{matsumura2014,
      title={Mitigation of the spectral dependent polarization angle response for achromatic half-wave plate}, 
      author={Tomotake Matsumura},
      year={2014},
      eprint={1404.5795},
      archivePrefix={arXiv},
      primaryClass={astro-ph.IM},
      url={https://arxiv.org/abs/1404.5795}, 
}

@misc{raffuzzi2026,
      title={Mitigating half-wave plate systematics at the map-making level: calibration requirements for LiteBIRD}, 
      author={N. Raffuzzi and A. Carones and M. Monelli and S. Giardiello and L. Pagano and Y. Sakurai and H. Ishino and E. Allys and A. Anand and J. Aumont and A. J. Banday and G. Barbieri Ripamonti and R. B. Barreiro and N. Bartolo and S. Basak and A. Basyrov and A. Besnard and M. Bortolami and T. Brinckmann and F. Cacciotti and E. Calabrese and P. Campeti and F. Carralot and F. J. Casas and J. Chandran and K. Cheung and M. Citran and L. Clermont and F. Columbro and A. Coppolecchia and F. Cuttaia and P. de Bernardis and T. de Haan and M. De Lucia and P. Diego-Palazuelos and H. K. Eriksen and F. Finelli and C. Franceschet and U. Fuskeland and G. Galloni and M. Galloway and M. Gerbino and M. Gervasi and T. Ghigna and C. Gimeno-Amo and A. Gruppuso and M. Hazumi and S. Henrot-Versillé and L. T. Hergt and E. Hivon and K. Kohri and L. Lamagna and M. Lattanzi and C. Leloup and F. Levrier and A. I. Lonappan and M. López-Caniego and G. Luzzi and J. Macias-Perez and V. Maranchery and E. Martínez-González and S. Masi and S. Matarrese and T. Matsumura and S. Micheli and M. Migliaccio and G. Morgante and L. Mousset and R. Nagata and T. Namikawa and P. Natoli and A. Novelli and F. Noviello and A. Occhiuzzi and A. Paiella and D. Paoletti and G. Pascual-Cisneros and G. Patanchon and F. Piacentini and G. Piccirilli and M. Pinchera and G. Polenta and L. Porcelli and M. Remazeilles and A. Ritacco and M. Ruiz-Granda and L. Salvati and J. Sanghavi and V. Sauvage and D. Scott and M. Shiraishi and G. Signorelli and R. M. Sullivan and Y. Takase and L. Terenzi and M. Tomasi and M. Tristram and L. Vacher and B. van Tent and P. Vielva and S. Vinzl and I. K. Wehus and G. Weymann-Despres and E. J. Wollack},
      year={2026},
      eprint={2602.02410},
      archivePrefix={arXiv},
      primaryClass={astro-ph.CO},
      url={https://arxiv.org/abs/2602.02410}, 
}

@misc{patanchon2023,
      title={Effect of Instrumental Polarization with a Half-Wave Plate on the {$B$}-Mode Signal: Prediction and Correction}, 
      author={Guillaume Patanchon and Hiroaki Imada and Hirokazu Ishino and Tomotake Matsumura},
      year={2023},
      eprint={2308.00967},
      archivePrefix={arXiv},
      primaryClass={astro-ph.CO},
      url={https://arxiv.org/abs/2308.00967}, 
}

@ARTICLE{yamada2024simons,
   author  = "K. Yamada and B. Bixler and Y. Sakurai and P. C. Ashton and J. Sugiyama and K. Arnold and J. Begin and L. Corbett and S. Day-Weiss and N. Galitzki and others",
   title   = "The {Simons Observatory}: Cryogenic half wave plate rotation mechanism for the small aperture telescopes",
   journal = "Rev. Sci. Instrum.",
   volume  = "95",
   number  = "2",
   year    = "2024"
}

@ARTICLE{sugiyama2024simons,
   author  = "J. Sugiyama and T. Terasaki and K. Sakaguri and B. Bixler and Y. Sakurai and K. Arnold and K. T. Crowley and R. Datta and N. Galitzki and M. Hasegawa and others",
   title   = "The {Simons Observatory}: Development and optical evaluation of achromatic half-wave plates",
   journal = "J. Low Temp. Phys.",
   volume  = "214",
   number  = "3",
   pages   = "173--181",
   year    = "2024"
}

@INPROCEEDINGS{hill2016design,
   author    = "C. A. Hill and S. Beckman and Y. Chinone and N. Goeckner-Wald and M. Hazumi and B. Keating and A. Kusaka and A. T. Lee and F. Matsuda and R. Plambeck and others",
   title     = "Design and development of an ambient-temperature continuously-rotating achromatic half-wave plate for {CMB} polarization modulation on the {{POLARBEAR}-2} experiment",
   booktitle = "Proc. SPIE Millim. Submillim. Far-Infrared Detect. Instrum. Astron. VIII",
   volume    = "9914",
   pages     = "699--716",
   year      = "2016"
}

@INPROCEEDINGS{klein2011cryogenic,
   author    = "J. Klein and A. Aboobaker and P. Ade and F. Aubin and C. Baccigalupi and C. Bao and J. Borrill and D. Chapman and J. Didier and M. Dobbs and others",
   title     = "A cryogenic half-wave plate polarimeter using a superconducting magnetic bearing",
   booktitle = "Proc. SPIE Cryogen. Opt. Syst. Instrum. XIII",
   volume    = "8150",
   pages     = "31--40",
   year      = "2011"
}

@ARTICLE{duivenvoorden2021probing,
   author  = "A. J. Duivenvoorden and A. E. Adler and M. Billi and N. Dachlythra and J. E. Gudmundsson",
   title   = "Probing frequency-dependent half-wave plate systematics for {CMB} experiments with full-sky beam convolution simulations",
   journal = "Mon. Not. R. Astron. Soc.",
   volume  = "502",
   number  = "3",
   pages   = "4526--4539",
   year    = "2021"
}

@ARTICLE{verges2021framework,
   author  = "C. Vergès and J. Errard and R. Stompor",
   title   = "Framework for analysis of next generation, polarized {CMB} data sets in the presence of Galactic foregrounds and systematic effects",
   journal = "Phys. Rev. D",
   volume  = "103",
   number  = "6",
   pages   = "063507",
   year    = "2021",
   doi     = "10.1103/PhysRevD.103.063507"
}

@ARTICLE{stompor2009maximum,
   author  = "R. Stompor and S. Leach and F. Stivoli and C. Baccigalupi",
   title   = "Maximum likelihood algorithm for parametric component separation in cosmic microwave background experiments",
   journal = "Mon. Not. R. Astron. Soc.",
   volume  = "392",
   number  = "1",
   pages   = "216--232",
   year    = "2009"
}

@ARTICLE{essinger2013transfer,
   author  = "T. Essinger-Hileman",
   title   = "Transfer matrix for treating stratified media including birefringent crystals",
   journal = "Appl. Opt.",
   volume  = "52",
   number  = "2",
   pages   = "212",
   year    = "2013"
}

@ARTICLE{poletti2017,
       author = {{Poletti}, Davide and {Fabbian}, Giulio and {Le Jeune}, Maude and {Peloton}, Julien and {Arnold}, Kam and {Baccigalupi}, Carlo and {Barron}, Darcy and {Beckman}, Shawn and {Borrill}, Julian and {Chapman}, Scott and et al.},
        title = "{Making maps of cosmic microwave background polarization for B-mode studies: the POLARBEAR example}",
      journal = {Astronomy and Astrophysics},
         year = 2017,
        month = apr,
       volume = {600},
          eid = {A60},
        pages = {A60},
          doi = {10.1051/0004-6361/201629467},
archivePrefix = {arXiv},
       eprint = {1608.01624},
 primaryClass = {astro-ph.IM},
       adsurl = {https://ui.adsabs.harvard.edu/abs/2017A&A...600A..60P}
}

@ARTICLE{stompor2016forecasting,
   author  = "R. Stompor and J. Errard and D. Poletti",
   title   = "Forecasting performance of {CMB} experiments in the presence of complex foreground contaminations",
   journal = "Phys. Rev. D",
   volume  = "94",
   number  = "8",
   pages   = "083526",
   year    = "2016"
}

@ARTICLE{sakaguri2024antireflection,
   author  = "K. Sakaguri and M. Hasegawa and Y. Sakurai and J. Sugiyama and N. Farias and C. A. Hill and B. R. Johnson and K. Konishi and A. Kusaka and A. T. Lee and T. Matsumura and E. J. Wollack and J. Yumoto",
   title   = "Anti-reflection coating with mullite and Duroid for large-diameter cryogenic sapphire and alumina optics",
   journal = "Appl. Opt.",
   volume  = "63",
   number  = "6",
   pages   = "1618",
   year    = "2024"
}

@ARTICLE{Kusaka_2014,
   title={Modulation of cosmic microwave background polarization with a warm rapidly rotating half-wave plate on the {Atacama B-Mode Search} instrument},
   volume={85},
   ISSN={1089-7623},
   url={http://dx.doi.org/10.1063/1.4862058},
   DOI={10.1063/1.4862058},
   number={2},
   journal={Review of Scientific Instruments},
   publisher={AIP Publishing},
   author={Kusaka, A. and Essinger-Hileman, T. and Appel, J. W. and Gallardo, P. and Irwin, K. D. and Jarosik, N. and Nolta, M. R. and Page, L. A. and Parker, L. P. and Raghunathan, S. and Sievers, J. L. and Simon, S. M. and Staggs, S. T. and Visnjic, K.},
   year={2014},
   month=feb }

@ARTICLE{Johnson_2007,
   title={MAXIPOL: Cosmic Microwave Background Polarimetry Using a Rotating Half‐Wave Plate},
   volume={665},
   ISSN={1538-4357},
   url={http://dx.doi.org/10.1086/518105},
   DOI={10.1086/518105},
   number={1},
   journal={The Astrophysical Journal},
   publisher={American Astronomical Society},
   author={Johnson, B. R. and Collins, J. and Abroe, M. E. and Ade, P. A. R. and Bock, J. and Borrill, J. and Boscaleri, A. and de Bernardis, P. and Hanany, S. and Jaffe, A. H. and Jones, T. and Lee, A. T. and Levinson, L. and Matsumura, T. and Rabii, B. and Renbarger, T. and Richards, P. L. and Smoot, G. F. and Stompor, R. and Tran, H. T. and Winant, C. D. and Wu, J. H. P. and Zuntz, J.},
   year={2007},
   month=aug, pages={42–54} }

@ARTICLE{El_Bouhargani_2022,
   title={MAPPRAISER: A massively parallel map-making framework for multi-kilo pixel {CMB} experiments},
   volume={39},
   ISSN={2213-1337},
   url={http://dx.doi.org/10.1016/j.ascom.2022.100576},
   DOI={10.1016/j.ascom.2022.100576},
   journal={Astronomy and Computing},
   publisher={Elsevier BV},
   author={El Bouhargani, H. and Jamal, A. and Beck, D. and Errard, J. and Grigori, L. and Stompor, R.},
   year={2022},
   month=apr, pages={100576} }

@misc{Kabalan2025,
      title={A GPU-Accelerated JAX Framework for Robust Parametric Component Separation and Clustering Optimization for CMB Polarization Satellites}, 
      author={Wassim Kabalan and Arianna Rizzieri and Wuhyun Sohn and Artem Basyrov and Alexandre Boucaud and Benjamin Beringue and Pierre Chanial and Ema Tsang King Sang and Josquin Errard},
      year={2026},
      eprint={2604.08463},
      archivePrefix={arXiv},
      primaryClass={astro-ph.CO},
      url={https://arxiv.org/abs/2604.08463}, 
}

@unpublished{Beringue2025,
  author       = {Beringue, B. and Villarrubia-Aguilar, A. and Goutaudier, E. and Errard, J.},
  title        = {Atmospheric Noise Modeling and Subtraction for Ground-based {CMB} Observations},
  note         = {in preparation},
  year         = {2026}
}

@unpublished{Errard2026,
  author       = {Errard, J. and {the SciPol collaboration}},
  note         = {in preparation},
  year         = {2027}
}

@BOOK{bryan2014half,
  title={Half-wave plates for the SPIDER cosmic microwave background polarimeter},
  author={Bryan, Sean Alan},
  year={2014},
  publisher={Case Western Reserve University}
}

@INPROCEEDINGS{Chanial2012,
       author = {{Chanial}, P. and {Barbey}, N.},
        title = "{PyOperators: Operators and solvers for high-performance computing}",
    booktitle = {SF2A-2012: Proceedings of the Annual meeting of the French Society of Astronomy and Astrophysics},
         year = 2012,
       editor = {{Boissier}, S. and {de Laverny}, P. and {Nardetto}, N. and {Samadi}, R. and {Valls-Gabaud}, D. and {Wozniak}, H.},
        month = dec,
        pages = {513-517},
       adsurl = {https://ui.adsabs.harvard.edu/abs/2012sf2a.conf..513C}
}

@article{Papez2020,
   title={Accelerating linear system solvers for time-domain component separation of cosmic microwave background data},
   volume={638},
   ISSN={1432-0746},
   url={http://dx.doi.org/10.1051/0004-6361/202037687},
   DOI={10.1051/0004-6361/202037687},
   journal={Astronomy \& Astrophysics},
   publisher={EDP Sciences},
   author={Pape\v{z}, J. and Grigori, L. and Stompor, R.},
   year={2020},
   month=jun, pages={A73} }

@misc{rader2023,
      title={Lineax: unified linear solves and linear least-squares in JAX and Equinox}, 
      author={Jason Rader and Terry Lyons and Patrick Kidger},
      year={2023},
      eprint={2311.17283},
      archivePrefix={arXiv},
      primaryClass={cs.MS},
      url={https://arxiv.org/abs/2311.17283}, 
}

@ARTICLE{Gorski_2005,
   title={HEALPix: A Framework for High‐Resolution Discretization and Fast Analysis of Data Distributed on the Sphere},
   volume={622},
   ISSN={1538-4357},
   url={http://dx.doi.org/10.1086/427976},
   DOI={10.1086/427976},
   number={2},
   journal={The Astrophysical Journal},
   publisher={American Astronomical Society},
   author={G\'orski, K. M. and Hivon, E. and Banday, A. J. and Wandelt, B. D. and Hansen, F. K. and Reinecke, M. and Bartelmann, M.},
   year={2005},
   month=apr, pages={759–771} }

@ARTICLE{Kamionkowski_2016,
   title={The Quest for B Modes from Inflationary Gravitational Waves},
   volume={54},
   ISSN={1545-4282},
   url={http://dx.doi.org/10.1146/annurev-astro-081915-023433},
   DOI={10.1146/annurev-astro-081915-023433},
   number={1},
   journal={Annual Review of Astronomy and Astrophysics},
   publisher={Annual Reviews},
   author={Kamionkowski, Marc and Kovetz, Ely D.},
   year={2016},
   month=sep, pages={227–269} }

@ARTICLE{Errard_2016,
   title={Robust forecasts on fundamental physics from the foreground-obscured, gravitationally-lensed {CMB} polarization},
   volume={2016},
   ISSN={1475-7516},
   url={http://dx.doi.org/10.1088/1475-7516/2016/03/052},
   DOI={10.1088/1475-7516/2016/03/052},
   number={03},
   journal={Journal of Cosmology and Astroparticle Physics},
   publisher={IOP Publishing},
   author={Errard, Josquin and Feeney, Stephen M. and Peiris, Hiranya V. and Jaffe, Andrew H.},
   year={2016},
   month=mar, pages={052–052} }

@ARTICLE{Takakura_2017,
   title={Performance of a continuously rotating half-wave plate on the {POLARBEAR} telescope},
   volume={2017},
   ISSN={1475-7516},
   url={http://dx.doi.org/10.1088/1475-7516/2017/05/008},
   DOI={10.1088/1475-7516/2017/05/008},
   number={05},
   journal={Journal of Cosmology and Astroparticle Physics},
   publisher={IOP Publishing},
   author={Takakura, Satoru and Aguilar, Mario and Akiba, Yoshiki and Arnold, Kam and Baccigalupi, Carlo and Barron, Darcy and Beckman, Shawn and Boettger, David and Borrill, Julian and Chapman, Scott and Chinone, Yuji and Cukierman, Ari and Ducout, Anne and Elleflot, Tucker and Errard, Josquin and Fabbian, Giulio and Fujino, Takuro and Galitzki, Nicholas and Goeckner-Wald, Neil and Halverson, Nils W. and Hasegawa, Masaya and Hattori, Kaori and Hazumi, Masashi and Hill, Charles and Howe, Logan and Inoue, Yuki and Jaffe, Andrew H. and Jeong, Oliver and Kaneko, Daisuke and Katayama, Nobuhiko and Keating, Brian and Keskitalo, Reijo and Kisner, Theodore and Krachmalnicoff, Nicoletta and Kusaka, Akito and Lee, Adrian T. and Leon, David and Lowry, Lindsay and Matsuda, Frederick and Matsumura, Tomotake and Navaroli, Martin and Nishino, Haruki and Paar, Hans and Peloton, Julien and Poletti, Davide and Puglisi, Giuseppe and Reichardt, Christian L. and Ross, Colin and Siritanasak, Praween and Suzuki, Aritoki and Tajima, Osamu and Takatori, Sayuri and Teply, Grant},
   year={2017},
   month=may, pages={008–008} }

@INPROCEEDINGS{pancharatnam1955,
  title={Achromatic combinations of birefringent plates: part I. An achromatic circular polarizer},
  author={Pancharatnam, Shivaramakrishnan},
  booktitle={Proceedings of the Indian Academy of Sciences-Section A},
  volume={41},
  number={4},
  pages={130--136},
  year={1955},
  organization={Springer}
}

@ARTICLE{Krachmalnicoff_2018,
   title={S–PASS view of polarized Galactic synchrotron at 2.3 GHz as a contaminant to {CMB} observations},
   volume={618},
   ISSN={1432-0746},
   url={http://dx.doi.org/10.1051/0004-6361/201832768},
   DOI={10.1051/0004-6361/201832768},
   journal = {Astronomy \& Astrophysics},
   publisher={EDP Sciences},
   author={Krachmalnicoff, N. and Carretti, E. and Baccigalupi, C. and Bernardi, G. and Brown, S. and Gaensler, B. M. and Haverkorn, M. and Kesteven, M. and Perrotta, F. and Poppi, S. and Staveley-Smith, L.},
   year={2018},
   month=oct, pages={A166} }

@ARTICLE{Delabrouille_2008,
   title={A full sky, low foreground, high resolution {CMB} map from {WMAP}},
   volume={493},
   ISSN={1432-0746},
   url={http://dx.doi.org/10.1051/0004-6361:200810514},
   DOI={10.1051/0004-6361:200810514},
   number={3},
   journal = {Astronomy \& Astrophysics},
   publisher={EDP Sciences},
   author={Delabrouille, J. and Cardoso, J.-F. and Le Jeune, M. and Betoule, M. and Fay, G. and Guilloux, F.},
   year={2008},
   month=nov, pages={835–857} }

@ARTICLE{Abitbol_2021,
   title={The {Simons Observatory}: gain, bandpass and polarization-angle calibration requirements for B-mode searches},
   volume={2021},
   ISSN={1475-7516},
   url={http://dx.doi.org/10.1088/1475-7516/2021/05/032},
   DOI={10.1088/1475-7516/2021/05/032},
   number={05},
   journal={Journal of Cosmology and Astroparticle Physics},
   publisher={IOP Publishing},
   author={Abitbol, Maximilian H. and Alonso, David and Simon, Sara M. and Lashner, Jack and Crowley, Kevin T. and Ali, Aamir M. and Azzoni, Susanna and Baccigalupi, Carlo and Barron, Darcy and Brown, Michael L. and Calabrese, Erminia and Carron, Julien and Chinone, Yuji and Chluba, Jens and Coppi, Gabriele and Crowley, Kevin D. and Devlin, Mark and Dunkley, Jo and Errard, Josquin and Fanfani, Valentina and Galitzki, Nicholas and Gerbino, Martina and Hill, J. Colin and Johnson, Bradley R. and Jost, Baptiste and Keating, Brian and Krachmalnicoff, Nicoletta and Kusaka, Akito and Lee, Adrian T. and Louis, Thibaut and Madhavacheril, Mathew S. and McCarrick, Heather and McMahon, Jeffrey and Meerburg, P. Daniel and Nati, Federico and Nishino, Haruki and Page, Lyman A. and Poletti, Davide and Puglisi, Giuseppe and Randall, Michael J. and Rotti, Aditya and Spisak, Jacob and Suzuki, Aritoki and Teply, Grant P. and Vergès, Clara and Wollack, Edward J. and Xu, Zhilei and Zannoni, Mario},
   year={2021},
   month=may, pages={032} }

@ARTICLE{Moncelsi_2013,
   title={Empirical modelling of the BLASTPol achromatic half-wave plate for precision submillimetre polarimetry},
   volume={437},
   ISSN={1365-2966},
   url={http://dx.doi.org/10.1093/mnras/stt2090},
   DOI={10.1093/mnras/stt2090},
   number={3},
   journal={Monthly Notices of the Royal Astronomical Society},
   publisher={Oxford University Press (OUP)},
   author={Moncelsi, Lorenzo and Ade, Peter A. R. and Angilè, Francesco E. and Benton, Steven J. and Devlin, Mark J. and Fissel, Laura M. and Gandilo, Natalie N. and Gundersen, Joshua O. and Matthews, Tristan G. and Netterfield, C. Barth and Novak, Giles and Nutter, David and Pascale, Enzo and Poidevin, Frédérick and Savini, Giorgio and Scott, Douglas and Soler, Juan Diego and Spencer, Locke D. and Truch, Matthew D. P. and Tucker, Gregory S. and Zhang, Jin},
   year={2013},
   month=nov, pages={2772–2789} }

@MISC{nakata2025,
      title={The {Simons Observatory}: Detector Polarization Angle Calibration using Sparse Wire Grid with Initial Data Sets of the Small Aperture Telescope}, 
      author={Hironobu Nakata and Shunsuke Adachi and Kyohei Yamada and Michael Randall and Yutaro Kasai and Kam Arnold and Bryce Bixler and Yuji Chinone and Kevin T. Crowley and Nadia Dachlythra and Samuel Day-Weiss and Nicholas Galitzki and Serena Giardiello and Bradley R. Johnson and Brian Keating and Brian J. Koopman and Akito Kusaka and Jack Lashner and Federico Nati and Lyman Page and Daichi Sasaki and Yoshinori Sueno and Junya Suzuki and Osamu Tajima and Tran Tsan},
      year={2025},
      eprint={2512.19102},
      archivePrefix={arXiv},
      primaryClass={astro-ph.IM},
      url={https://arxiv.org/abs/2512.19102}, 
}

@ARTICLE{Coppi_2025,
   title={PROTOCALC, a W-band Polarized Calibrator for Cosmic Microwave Background Telescopes: Application to {Simons Observatory} and CLASS},
   volume={279},
   ISSN={1538-4365},
   url={http://dx.doi.org/10.3847/1538-4365/adde5f},
   DOI={10.3847/1538-4365/adde5f},
   number={1},
   journal={The Astrophysical Journal Supplement Series},
   publisher={American Astronomical Society},
   author={Coppi, Gabriele and Dachlythra, Nadia and Nati, Federico and Dünner-Planella, Rolando and Adler, Alexandre E. and Errard, Josquin and Galitzki, Nicholas and Li, Yunyang and Petroff, Matthew A. and Simon, Sara M. and Sang, Ema Tsang King and Aguilar, Amalia Villarrubia and Wollack, Edward J. and Zannoni, Mario},
   year={2025},
   month=jul, pages={30} }

@ARTICLE{Wolz_2024,
   title={The {Simons Observatory}: Pipeline comparison and validation for large-scale B-modes},
   volume={686},
   ISSN={1432-0746},
   url={http://dx.doi.org/10.1051/0004-6361/202346105},
   DOI={10.1051/0004-6361/202346105},
   journal={Astronomy \& Astrophysics},
   publisher={EDP Sciences},
   author={Wolz, Kevin and Azzoni, Susanna and Hervías-Caimapo, Carlos and Errard, Josquin and Krachmalnicoff, Nicoletta and Alonso, David and Baccigalupi, Carlo and Baleato Lizancos, Antón and Brown, Michael L. and Calabrese, Erminia and Chluba, Jens and Dunkley, Jo and Fabbian, Giulio and Galitzki, Nicholas and Jost, Baptiste and Morshed, Magdy and Nati, Federico},
   year={2024},
   month=may, pages={A16} }

@MISC{hertig2024,
      title={The {Simons Observatory}: Combining delensing and foreground cleaning for improved constraints on inflation}, 
      author={Emilie Hertig and Kevin Wolz and Toshiya Namikawa and Antón Baleato Lizancos and Susanna Azzoni and Anthony Challinor},
      year={2024},
      eprint={2405.13201},
      archivePrefix={arXiv},
      primaryClass={astro-ph.CO},
      url={https://arxiv.org/abs/2405.13201}, 
}

@ARTICLE{Monelli_2023,
   title={Impact of half-wave plate systematics on the measurement of cosmic birefringence from {CMB} polarization},
   volume={2023},
   ISSN={1475-7516},
   url={http://dx.doi.org/10.1088/1475-7516/2023/03/034},
   DOI={10.1088/1475-7516/2023/03/034},
   number={03},
   journal={Journal of Cosmology and Astroparticle Physics},
   publisher={IOP Publishing},
   author={Monelli, Marta and Komatsu, Eiichiro and Adler, Alexandre E. and Billi, Matteo and Campeti, Paolo and Dachlythra, Nadia and Duivenvoorden, Adriaan J. and Gudmundsson, Jon E. and Reinecke, Martin},
   year={2023},
   month=mar, pages={034} }

@ARTICLE{Essinger_Hileman_2016,
   title={Systematic effects from an ambient-temperature, continuously rotating half-wave plate},
   volume={87},
   ISSN={1089-7623},
   url={http://dx.doi.org/10.1063/1.4962023},
   DOI={10.1063/1.4962023},
   number={9},
   journal={Review of Scientific Instruments},
   publisher={AIP Publishing},
   author={Essinger-Hileman, T. and Kusaka, A. and Appel, J. W. and Choi, S. K. and Crowley, K. and Ho, S. P. and Jarosik, N. and Page, L. A. and Parker, L. P. and Raghunathan, S. and Simon, S. M. and Staggs, S. T. and Visnjic, K.},
   year={2016},
   month=sep }

@MISC{salatino2018,
      title={Studies of Systematic Uncertainties for {Simons Observatory}: Polarization Modulator Related Effects}, 
      author={Maria Salatino and Jacob Lashner and Martina Gerbino and Sara M. Simon and Joy Didier and Aamir Ali and Peter C. Ashton and Sean Bryan and Yuji Chinone and Kevin Coughlin and Kevin T. Crowley and Giulio Fabbian and Nicholas Galitzki and Neil Goeckner-Wald and Joseph E. Golec and Jon E. Gudmundsson and Charles A. Hill and Brian Keating and Akito Kusaka and Adrian T. Lee and Jeffrey McMahon and Amber D. Miller and Giuseppe Puglisi and Christian L. Reichardt and Grant Teply and Zhilei Xu and Ningfeng Zhu},
      year={2018},
      eprint={1808.07442},
      archivePrefix={arXiv},
      primaryClass={astro-ph.IM},
      url={https://arxiv.org/abs/1808.07442}, 
}

@ARTICLE{Thornton_2016,
   title={THE ATACAMA COSMOLOGY TELESCOPE: THE POLARIZATION-SENSITIVE ACTPol INSTRUMENT},
   volume={227},
   ISSN={1538-4365},
   url={http://dx.doi.org/10.3847/1538-4365/227/2/21},
   DOI={10.3847/1538-4365/227/2/21},
   number={2},
   journal={The Astrophysical Journal Supplement Series},
   publisher={American Astronomical Society},
   author={Thornton, R. J. and Ade, P. A. R. and Aiola, S. and Angilè, F. E. and Amiri, M. and Beall, J. A. and Becker, D. T. and Cho, H-M. and Choi, S. K. and Corlies, P. and Coughlin, K. P. and Datta, R. and Devlin, M. J. and Dicker, S. R. and Dünner, R. and Fowler, J. W. and Fox, A. E. and Gallardo, P. A. and Gao, J. and Grace, E. and Halpern, M. and Hasselfield, M. and Henderson, S. W. and Hilton, G. C. and Hincks, A. D. and Ho, S. P. and Hubmayr, J. and Irwin, K. D. and Klein, J. and Koopman, B. and Li, Dale and Louis, T. and Lungu, M. and Maurin, L. and McMahon, J. and Munson, C. D. and Naess, S. and Nati, F. and Newburgh, L. and Nibarger, J. and Niemack, M. D. and Niraula, P. and Nolta, M. R. and Page, L. A. and Pappas, C. G. and Schillaci, A. and Schmitt, B. L. and Sehgal, N. and Sievers, J. L. and Simon, S. M. and Staggs, S. T. and Tucker, C. and Uehara, M. and Lanen, J. van and Ward, J. T. and Wollack, E. J.},
   year={2016},
   month=dec, pages={21} }

@ARTICLE{Ward_2018,
   title={The Effects of Bandpass Variations on Foreground Removal Forecasts for Future {CMB} Experiments},
   volume={861},
   ISSN={1538-4357},
   url={http://dx.doi.org/10.3847/1538-4357/aac71f},
   DOI={10.3847/1538-4357/aac71f},
   number={2},
   journal={The Astrophysical Journal},
   publisher={American Astronomical Society},
   author={Ward, J. T. and Alonso, D. and Errard, J. and Devlin, M. J. and Hasselfield, M.},
   year={2018},
   month=jul, pages={82} }

@ARTICLE{Planck2020,
   title={Planck
                    2018 results: VI. Cosmological parameters},
   volume={641},
   ISSN={1432-0746},
   url={http://dx.doi.org/10.1051/0004-6361/201833910},
   DOI={10.1051/0004-6361/201833910},
   journal={Astronomy \& Astrophysics},
   publisher={EDP Sciences},
   author={Aghanim, N. and Akrami, Y. and Ashdown, M. and Aumont, J. and Baccigalupi, C. and Ballardini, M. and Banday, A. J. and Barreiro, R. B. and Bartolo, N. and Basak, S. and Battye, R. and Benabed, K. and Bernard, J.-P. and Bersanelli, M. and Bielewicz, P. and Bock, J. J. and Bond, J. R. and Borrill, J. and Bouchet, F. R. and Boulanger, F. and Bucher, M. and Burigana, C. and Butler, R. C. and Calabrese, E. and Cardoso, J.-F. and Carron, J. and Challinor, A. and Chiang, H. C. and Chluba, J. and Colombo, L. P. L. and Combet, C. and Contreras, D. and Crill, B. P. and Cuttaia, F. and de Bernardis, P. and de Zotti, G. and Delabrouille, J. and Delouis, J.-M. and Di Valentino, E. and Diego, J. M. and Doré, O. and Douspis, M. and Ducout, A. and Dupac, X. and Dusini, S. and Efstathiou, G. and Elsner, F. and Enßlin, T. A. and Eriksen, H. K. and Fantaye, Y. and Farhang, M. and Fergusson, J. and Fernandez-Cobos, R. and Finelli, F. and Forastieri, F. and Frailis, M. and Fraisse, A. A. and Franceschi, E. and Frolov, A. and Galeotta, S. and Galli, S. and Ganga, K. and Génova-Santos, R. T. and Gerbino, M. and Ghosh, T. and González-Nuevo, J. and Górski, K. M. and Gratton, S. and Gruppuso, A. and Gudmundsson, J. E. and Hamann, J. and Handley, W. and Hansen, F. K. and Herranz, D. and Hildebrandt, S. R. and Hivon, E. and Huang, Z. and Jaffe, A. H. and Jones, W. C. and Karakci, A. and Keihänen, E. and Keskitalo, R. and Kiiveri, K. and Kim, J. and Kisner, T. S. and Knox, L. and Krachmalnicoff, N. and Kunz, M. and Kurki-Suonio, H. and Lagache, G. and Lamarre, J.-M. and Lasenby, A. and Lattanzi, M. and Lawrence, C. R. and Le Jeune, M. and Lemos, P. and Lesgourgues, J. and Levrier, F. and Lewis, A. and Liguori, M. and Lilje, P. B. and Lilley, M. and Lindholm, V. and López-Caniego, M. and Lubin, P. M. and Ma, Y.-Z. and Macías-Pérez, J. F. and Maggio, G. and Maino, D. and Mandolesi, N. and Mangilli, A. and Marcos-Caballero, A. and Maris, M. and Martin, P. G. and Martinelli, M. and Martínez-González, E. and Matarrese, S. and Mauri, N. and McEwen, J. D. and Meinhold, P. R. and Melchiorri, A. and Mennella, A. and Migliaccio, M. and Millea, M. and Mitra, S. and Miville-Deschênes, M.-A. and Molinari, D. and Montier, L. and Morgante, G. and Moss, A. and Natoli, P. and Nørgaard-Nielsen, H. U. and Pagano, L. and Paoletti, D. and Partridge, B. and Patanchon, G. and Peiris, H. V. and Perrotta, F. and Pettorino, V. and Piacentini, F. and Polastri, L. and Polenta, G. and Puget, J.-L. and Rachen, J. P. and Reinecke, M. and Remazeilles, M. and Renzi, A. and Rocha, G. and Rosset, C. and Roudier, G. and Rubiño-Martín, J. A. and Ruiz-Granados, B. and Salvati, L. and Sandri, M. and Savelainen, M. and Scott, D. and Shellard, E. P. S. and Sirignano, C. and Sirri, G. and Spencer, L. D. and Sunyaev, R. and Suur-Uski, A.-S. and Tauber, J. A. and Tavagnacco, D. and Tenti, M. and Toffolatti, L. and Tomasi, M. and Trombetti, T. and Valenziano, L. and Valiviita, J. and Van Tent, B. and Vibert, L. and Vielva, P. and Villa, F. and Vittorio, N. and Wandelt, B. D. and Wehus, I. K. and White, M. and White, S. D. M. and Zacchei, A. and Zonca, A.},
   year={2020},
   month=sep, pages={A6} }

@misc{chanial2026,
      title={Furax: A Modular JAX Framework for Linear Operators in Astrophysical and Cosmological Data Analysis}, 
      author={Pierre Chanial and Simon Biquard and Wassim Kabalan and Wuhyun Sohn and Artem Basyrov and Benjamin Beringue and Alexandre Boucaud and Andréa Landais and Magdy Morshed and Radek Stompor and Ema Tsang King Sang and Amalia Villarrubia-Aguilar and Josquin Errard},
      year={2026},
      eprint={2603.19600},
      archivePrefix={arXiv},
      primaryClass={astro-ph.IM},
      url={https://arxiv.org/abs/2603.19600}, 
}

@unpublished{Sohn2026,
  author       = {Sohn, W. and et al.},
  title        = {Robust {CMB} polarisation mapmaking with a rotating half wave plate},
  note         = {in preparation},
  year         = {2026}
}

\end{document}